\documentclass[twocolumn]{aa}  

\usepackage{graphicx}
\usepackage{txfonts}
\usepackage{hyperref}
\usepackage{bm}
\usepackage{hyperref}
\usepackage[normalem]{ulem}
\usepackage[dvipsnames]{xcolor}
\newcommand{\refig}[1]{Fig.~\ref{#1}}
\newcommand{\refeq}[1]{Eq.~\ref{#1}}
\newcommand{\pluto}[0]{\texttt{PLUTO}}

\newcommand{\rmhd}[3]{\texttt{s#1e#2r#3}}
\newcommand{\pic}[1]{\texttt{s#1pic}}

\graphicspath{fig}
\begin{document}

   \title{Relativistic reconnection with effective resistivity}

   \subtitle{I. Dynamics and reconnection rate}

   \author{
        M. Bugli        \inst{1,2,3}
        \and
        E. F. Lopresti    \inst{1}
        \and
        E. Figueiredo   \inst{4}
        \and
        A. Mignone      \inst{1,3}
        \and
        B. Cerutti      \inst{4}
        \and
        G. Mattia       \inst{5,6}
        \and
        L. Del Zanna    \inst{7,6,8}
        \and\\
        G. Bodo         \inst{9}
        \and 
        V. Berta        \inst{1}
          }

    \institute{
    Dipartimento di Fisica, Universit\`a di Torino, Via P. Giuria 1, I-10125 Torino, Italy\\
    \email{matteo.bugli@unito.it}
    \and
    Universit\'e Paris-Saclay, Universit\'e Paris Cit\'e, CEA, CNRS, AIM, F-91191, Gif-sur-Yvette, France
    \and
    INFN - sezione di Torino , Via Pietro Giuria 1, I-10125 Torino, Italy
    \and                
    Univ. Grenoble Alpes, CNRS, IPAG, 38000 Grenoble, France
    \and
    Max-Planck Institute for Astronomy (MPIA), K{\"o}nigstuhl 17, 69117 Heidelberg, Germany
    \and
    INFN , Sezione di Firenze, Via G. Sansone 1, I-50019 Sesto Fiorentino (FI), Italy
    \and  
    Dipartimento di Fisica e Astronomia, Universit\`a di Firenze, Via G. Sansone 1, I-50019 Sesto Fiorentino (FI), Italy
    \and           
    INAF, Osservatorio Astrofisico di Arcetri, Largo E. Fermi 5, I-50125 Firenze, Italy
    \and
    INAF, Osservatorio Astrofisico di Torino, Strada Osservatorio 20, I-10025 Pino Torinese (TO), Italy
    }

    \date{Received September 15, 1996; accepted March 16, 1997}
    \date{}
    
    \abstract
    % context heading
    {Relativistic magnetic reconnection is one of the most fundamental mechanisms considered responsible for the acceleration of relativistic particles in astrophysical jets and magnetospheres of compact objects.
    Understanding the properties of the dissipation of magnetic fields and the formation of non-ideal electric fields is of paramount importance to quantify the efficiency of reconnection at energizing charged particles.}
    % aims heading
    {Recent results from particle-in-cell (PIC) simulations suggest that the fundamental properties of how magnetic fields dissipate in a current sheet might be captured by an ``effective resistivity'' formulation, which would locally enhance the amount of magnetic energy dissipated and favor the onset of fast reconnection.
    Our goal is to assess this ansatz quantitatively by comparing fluid models of magnetic reconnection with a non-constant magnetic diffusivity and fully-kinetic models.}
    % methods heading
    {We perform 2D resistive relativistic magnetohydrodynamic (ResRMHD) simulations of magnetic reconnection combined to PIC simulations using the same initial conditions (namely a Harris current sheet).
    We explore the impact of crucial parameters such as the plasma magnetization, its mass density, the grid resolution, and the characteristic plasma skin depth.
    }
    % results heading
    {Our ResRMHD models with effective resistivity can quantitatively reproduce the dynamics of fully-kinetic models of relativistic magnetic reconnection.
    In particular, they lead to reconnection rates consistent with PIC simulations, while for constant-resistivity fluid models the reconnection dynamics is generally 10 times slower.
    Even at modest resolutions the adoption of an effective resistivity can qualitatively capture the properties of kinetic reconnection models and produce reconnection rates compatible with collisionless models, i.e. of the order of $\sim10^{-1}$.}
    % conclusions heading (optional)
    {}

    \keywords{
                magnetic reconnection --
                magnetohydrodynamics (MHD) --
                relativistic processes -- 
                acceleration of particles --
                methods: numerical -- 
                plasmas                
               }

    \maketitle

%%%%%%%%%%%%%%%%%%%%%%%%%%%%%%%%%%%%%%%%%%%%%%%%%%%%%%%%%%%%%%%%%%%%%%%%%%%%%%%%
%%%%%%%%%%%%%%%%%%%%%%%%%%%%%%%%%%%%%%%%%%%%%%%%%%%%%%%%%%%%%%%%%%%%%%%%%%%%%%%%
%%%%%%%%%%%%%%%%%%%%%%%%%%%%%%%%%%%%%%%%%%%%%%%%%%%%%%%%%%%%%%%%%%%%%%%%%%%%%%%%
%%%%%%%%%%%%%%%%%%%%%%%%%%%%%%%%%%%%%%%%%%%%%%%%%%%%%%%%%%%%%%%%%%%%%%%%%%%%%%%%
%%%%%%%%%%%%%%%%%%%%%%%%%%%%%%%%%%%%%%%%%%%%%%%%%%%%%%%%%%%%%%%%%%%%%%%%%%%%%%%%

    \section{Introduction}
% Astrophysical context
Relativistic magnetic reconnection is one of the main physical processes invoked to explain the high-energy emission produced by strongly magnetized astrophysical plasmas.
Its core mechanism consists in the rearrangement of magnetic field lines, which rapidly transfers the magnetic energy stored in the plasma into heat, high-speed flows, and accelerated particles.
This process is believed to power a large number of highly energetic peculiar transients, such as flares from black hole magnetospheres \citep{nathanail2021,ripperda2022}, pulsar magnetospheres \citep{lyubarskii1996, uzdensky2014} and pulsar wind nebulae \citep{cerutti2012a,olmi2016}, fast radio bursts \cite{mahlmann2022,most2023}, gamma-ray bursts \cite{zhang2011,mckinney2012}, and flares from blazar jets \citep{giannios2013,petropoulou2016}.

% PIC vs. ResRMHD
Particle in cell (PIC) numerical simulations are an essential numerical tool to model the fundamental physics behind the reconnection of current sheets in astrophysical plasmas \citep{zenitani2001,cerutti2012b,sironi2014}.
Their ability to capture the collisionless nature of astrophysical plasmas allows one to quantitatively model the dissipation of magnetic fields and the consequent acceleration of relativistic particles from first principles.
However, their high computational cost poses a serious challenge when trying to address the large separation between typical length scales of the plasma (e.g. particles gyroradius, plasma skin depth) and the size of the astrophysical systems where reconnection takes place. 
On the other hand, resistive relativistic magnetohydrodynamics \citep[ResRMHD; see, e.g., ][]{komissarov2007,palenzuela2009} offers a powerful framework to study the impact of magnetic dissipation on the large-scale dynamics of accretion flows around compact objects \citep{qian2017,ripperda2019b,vos2024} and the launch of relativistic polar outflows \citep{mattia2023,mattia2024}.
By introducing an explicit physical magnetic diffusivity, such simulations can quantitatively address the formation of current sheets and the departure from an ideal relativistic magnetohydrodynamic (RMHD) regime \citep{delzanna2016}, while also decoupling the dissipation properties from the specific numerical implementation of the code adopted. 
Nonetheless, such resistivity is only a proxy for the actual magnetic dissipation occurring in the plasma, as its value is typically assumed to be constant across the domain and is significantly limited by the numerical diffusion imposed by the grid discretization.
Moreover, reconnection rates produced by fluid models are generally one order of magnitude lower than those obtained with fully-kinetic models \citep{sironi2014}, suggesting a fundamental shortcoming in the ability of ResRMHD simulations to capture the properties of collisionless reconnection.  

% Effective resistivity (Ripperda et al. 2019, ...) 
A possible way to improve the description of magnetic dissipation in the ResRMHD framework is to introduce an ``anomalous resistivity'' in specific regions in proximity of the current sheet, where the dissipation is significantly enhanced \cite{watanabe2006,zanotti2011}.
Such approach can significantly increase the rate at magnetic field lines are reconnected, but it requires to select a-priori a fraction of the numerical domain where dissipation needs to be amplified.
Some works adopted, instead, the more practical strategy to increase resistivity in regions with high values of current density \cite{zenitani2010,ripperda2019a}, which is more flexible and can deal with non-trivial configurations of the current sheet.
This prescription can lead to faster reconnection as well, but the amplitude of the ``effective'' resistivity was still set by an arbitrary constant that was not constrained by a specific physical model of collisionless reconnection.
% Selvi et al. 2023
\cite{selvi2023} provided the first physically-motivated model for effective resistivity informed by first-principles PIC simulations of 2D Harris sheets that is consistent with a classical collisionless inertial resistivity associated to a finite permanence of charged particles within the reconnecting region \citep{speiser1970}.
By measuring the dominant component of the electric field within the reconnecting layer in the comoving frame and comparing it to the local current density, they constructed a closed form for an effective resistivity $\eta_{\rm eff}$ that depends on the particles mean velocity and density.

% This paper
The present study aims at testing, for the first time, the prescription proposed by \cite{selvi2023} by using 2D ResRMHD models of magnetic reconnection.
We reformulated the expression for the effective resistivity in terms of fluid quantities available in MHD models and measured its effects on the dynamics of a standard 2D Harris current sheet, quantifying the impact of the different quantities that set its magnitude.
By performing a set of PIC simulations sharing the same initial conditions as the fluid models, we quantified the accuracy of the effective resistivity prescription to capture the non-collisional properties of magnetic dissipation, obtaining very good agreement between the two numerical frameworks.

Our study is presented following this structure.
In Section~\ref{sec:methods} we introduce the numerical methods we adopted and the derivation of the closed form of $\eta_{\rm eff}$ we used in the rest of the paper.
We then present in Section~\ref{sec:analysis} a detailed analysis of the reconnection dynamics introduced by the effective resistivity and the comparison with the constant resistivity and kinetic cases.
In Section~\ref{sec:parameters} we quantify instead the impact of numerical dissipation (through a convergence study) and physical parameters (such as magnetization, density, plasma skin depth) in combination with PIC models.
Finally, we share our conclusions in Section~\ref{sec:conclusions} and present numerical benchmarks validating our resistivity implementation in Appendix A.
We leave the analysis of the particle acceleration process and the impact of effective resistivity on their spectral distribution to an upcoming follow-up study.
CGS Gaussian units will be used throughout the paper.

%%%%%%%%%%%%%%%%%%%%%%%%%%%%%%%%%%%%%%%%%%%%%%%%%%%%%%%%%%%%%%%%%%%%%%%%%%%%%%%%
%%%%%%%%%%%%%%%%%%%%%%%%%%%%%%%%%%%%%%%%%%%%%%%%%%%%%%%%%%%%%%%%%%%%%%%%%%%%%%%%
%%%%%%%%%%%%%%%%%%%%%%%%%%%%%%%%%%%%%%%%%%%%%%%%%%%%%%%%%%%%%%%%%%%%%%%%%%%%%%%%
%%%%%%%%%%%%%%%%%%%%%%%%%%%%%%%%%%%%%%%%%%%%%%%%%%%%%%%%%%%%%%%%%%%%%%%%%%%%%%%%
%%%%%%%%%%%%%%%%%%%%%%%%%%%%%%%%%%%%%%%%%%%%%%%%%%%%%%%%%%%%%%%%%%%%%%%%%%%%%%%%

\section{Numerical methods}\label{sec:methods}

\subsection{Resistive RMHD models}
Our ResRMHD simulations are performed using the \pluto{} code \cite{mignone2007}.
We solve a set of conservation laws in the form
\begin{equation}\label{eq:cons_laws}
 \partial_t U = -\vec{\nabla}\cdot\tens{F}(U) + S(U) 
\end{equation}
with the vectors of conservative variables $U$ and fluxes $\tens{F}$ given by
\begin{equation}\label{eq:ResRMHD}
U = \left( \begin{array}{c}
D \\
m_i \\
{\cal E} \\
B_i \\
E_i 
\end{array}\right), \quad
 \tens{F}_j = \left( \begin{array}{c}
  \rho u_j \\
  T_{ij} \\
  c^2 m_j \\
  c \varepsilon^{ijk} E_k \\
  - c \varepsilon^{ijk} B_k
 \end{array}\right) , 
\end{equation}
where $D=\Gamma\rho$ is the mass density of the plasma, $\vec{m}= \rho h \Gamma\,\vec{u} + \tfrac{1}{4\pi c}\vec{E}\times\vec{B}$ is the total momentum density, ${\cal E}=\rho c^2 h\Gamma^2 - p + u_{\rm em}$ is the total energy density, $T_{ij}=\rho h u_i u_j -\tfrac{1}{4\pi}(E_i E_j + B_i B_j) + p_\mathrm{tot} \delta_{ij}$ is the total stress tensor, $\vec{B}$ is the magnetic field, $\vec{E}$ is the electric field (all measured in the laboratory frame), and $\varepsilon^{ijk}$ is the 3D Levi-Civita symbol.
In the previous expressions we introduced the Lorentz factor $\Gamma$, the electromagnetic energy density $u_{\rm em} = \tfrac{1}{8\pi} (E^2 + B^2)$, the total pressure $p_{\rm tot}=p+u_{\rm em}$, the specific enthalpy $h$, and the so-called ``primitive'' variables, i.e. the rest mass density $\rho$, the fluid spatial velocity $\vec{v}$, or the related spatial part of the four-velocity $\vec{u}=\Gamma\vec{v}$ (then $\Gamma = \sqrt{1+(u/c)^2}$), and the thermal pressure $p$.

The source term vector in \refeq{eq:cons_laws} is instead given by
\begin{equation} \label{eq:sources}
  S = \left( \begin{array}{c}
    0_{\times8} \\
    - 4\pi J_i
   \end{array}\right) ,
\end{equation}
where the current density $\vec{J}$ is expressed in terms of primitive variables and the electromagnetic field using the relativistic Ohm's law for a resistive plasma \citep{komissarov2007}
\begin{equation}\label{eq:Etilde}
    \vec{J} = \rho_e\vec{v} + \frac{1}{\eta} \Gamma \left[\,\vec{E} + \frac{1}{c}\vec{v}\times\vec{B} - \frac{1}{c^2}(\vec{E}\cdot\vec{v})\vec{v}\,\right]
\end{equation}
where $\rho_e$ is the electric charge density and $\eta$ is the electric resistivity, taken here as the inverse of the fluid conductivity.

A well-known property of the evolution equation for the electric field, as first pointed by \cite{komissarov2007}, is to be potentially stiff for low values of the magnetic resistivity, since parts of its source term (i.e. those inversely proportional to $\eta$) change over characteristic time-scales that can become much smaller than the rest of the contributions\citep{palenzuela2009,bucciantini2013,delzanna2018}.  
To ensure numerical stability to the time integration of Eqs.~\ref{eq:ResRMHD} we use the IMEX-SSP3(3,3,2) scheme \citep{pareschi2005}, following a similar procedure to the one presented in \cite{mignone2019} and \cite{tomei2020}.

The evolutionary laws described in Eq.~\ref{eq:cons_laws}, \ref{eq:ResRMHD}, and \ref{eq:sources} need to be coupled to the solenoidal constraint for the magnetic field and the conservation of electric charge, i.e.
\begin{align}
    &\vec{\nabla}\cdot\vec{B} = 0\; , \\
    &\vec{\nabla}\cdot\vec{E} = 4\pi\rho_e\; .
\end{align}
While we enforce the diverge-free condition by using staggered magnetic field components via the Constrained Transport (CT) method \citep{londrillo2004}, we use the latter equation to compute the electric charge without enforcing the electric field divergence \citep{bucciantini2013}.
All our models were performed using also WENOZ \citep{castro2011} reconstruction and the MHLLC Riemann solver \citep{mignone2019}. 
\begin{table}
    \centering
    \caption{List of ResRMHD and PIC models and their parameters: upstream ``cold'' magnetization, Alfv\`en speed, resistivity, effective plasma skin depth, upstream density, and magnetic field.}
    \label{tab:models}
    \renewcommand{\arraystretch}{1.2} % Default value: 1
    \begin{tabular}{c c c c c c c}
        \hline
        Model & $\sigma_0$ & $c_{A,0}$ & $\bar{\eta}$ & $\bar{\delta}_u$ & $\rho_0$ & $B_0$ \\
        \hline
        \hline
        \rmhd{10}{E}{1}   &  10 & 0.945 & $\eta_{\rm eff}$ & 0.002 & 1 & 3.16 \\
        \rmhd{10}{Ed0}{1}   &  10 & 0.945 & $\eta_{\rm eff}$ & 0.02 & 1 & 3.16 \\
        \rmhd{10}{C2}{1}   &  10 & 0.945 & $2.5\times10^{-2}$  & - & 1 & 3.16\\
        \rmhd{10}{C3}{1}   &  10 & 0.945 & $2.5\times10^{-3}$  & - & 1 & 3.16\\
        \rmhd{10}{C4}{1}   &  10 & 0.945 & $2.5\times10^{-4}$  & - & 1 & 3.16\\
        \rmhd{10}{C5}{1}   &  10 & 0.945 & $2.5\times10^{-5}$  & - & 1 & 3.16\\
        \rmhd{10}{E}{01}  &  10 & 0.945 & $\eta_{\rm eff}$ & 0.002 & 0.1 & 1 \\
        \rmhd{10}{E}{10}  &  10 & 0.945 & $\eta_{\rm eff}$ & 0.002 & 10 & 10\\
        \rmhd{4}{E}{1}    &  4  & 0.887 & $\eta_{\rm eff}$ & 0.002 & 1 & 2 \\      
        \rmhd{1}{E}{1}    &  1  & 0.704 & $\eta_{\rm eff}$ & 0.002 & 1 & 1 \\
        \hline
        \pic{10}          &  10 & 0.945 & - & - & - & - \\
        \pic{4}           &  4  & 0.887 & - & - & - & - \\
        \pic{1}           &  1  & 0.704 & - & - & - & - \\
        \hline
        \hline
    \end{tabular}
\end{table}
%
%-------------------------------------------------------------------------------
%
\subsubsection{Effective resistivity model}
We now describe our implementation in the \pluto{} code of the effective resistivity model proposed by \cite{selvi2023}, which applies exclusively to a pair plasma without guide field.
Using 2D fully-kinetic relativistic PIC models of reconnecting Harris sheets, \cite{selvi2023} measured the resistive component of the electric field produced within the current sheet and compared it to the general expression for the relativistic Ohm law \cite{hesse2007}.
In their model the sheet is aligned with the $x$-axis, the $y$ direction is transverse to it, and the $z$-axis is a symmetry axis perpendicular to the plane containing the initial magnetic field (i.e. no guide field is included). 
They identified the effective resistivity parameter as the ratio between the out-of-plane components of the comoving electric field and the current density and then estimated the dominant terms contributing to the electric field in the co-moving frame of the fluid (which is expected to vanish only in the limit of a perfectly conducting plasma), obtaining
\begin{equation}\label{eq:eff_eta1}
     \eta_{\rm eff} \simeq \frac{m}{n_t e^2} \frac{ \langle u_{\alpha z} \rangle}{ \langle v_{\alpha z} \rangle} \partial_y \langle v_{\alpha y} \rangle,
\end{equation}
where $m$ is the mass of the electron, $n_t$ is the plasma number density, $e$ is the electron charge, $u_{\alpha z}$ is the $z-$component of the four-velocity of the species $\alpha$ (in our case, either an electron or a positron), $v_{\alpha z}$ is the $z-$component of its three-velocity, and $\langle\rangle$ represents an average over the particle distribution.
It should be noted that \refeq{eq:eff_eta1} is not covariant, and its validity is limited, a priori, to the fluid comoving frame \citep[given the expression of the Ohm law used by][]{selvi2023}.
While a covariant formulation would be much preferable, in the absence of fast bulk motions (such as in our problem) there should not be strong deviations in the effective resistivity.
To express the effective resistivity exclusively in terms of fluid quantities that are available in an RMHD simulation, one can first approximate the ratio $\langle u_{\alpha z} \rangle/\langle v_{\alpha z}\rangle$ with the electron Lorentz factor $\Gamma_e$ and introduce the fluid rest-mass density $\rho=mn_t$ and the $y-$component of the fluid velocity $v_y\approx\langle v_{\alpha y} \rangle$ (using charge neutrality and symmetry reasons).
If we now express $\Gamma_e$ in terms of the current density (i.e. $\rho e\langle v_{ei}\rangle / m\simeq j_i$) we can use the fact that for a 2D current sheet with no guide field the only non-vanishing component of the current density is the out-of-plane one and obtain 
\begin{equation}\label{eq:eff_eta3}
     \eta_{\rm eff} = \frac{mc}{e} \frac{\partial_y v_y}{\sqrt{(\rho ec/m)^2 - j_z^2}}\; .
\end{equation}
This relation corresponds to a form of the effective resistivity for a fluid code proposed in an earlier version of \cite{selvi2023} and can be rewritten as
\begin{equation}
    \eta_{\rm eff} = \frac{E_0}{\sqrt{J_0^2 - j_z^2}}\; ,
\end{equation}
where we defined the quantities $E_0 = (mc/e)\partial_y v_y$ and $J_0 = \rho ec/m$.
By substituting $j_z=e_z/\eta_\mathrm{eff}$ and solving for $\eta_{\rm eff}$, we finally obtain
\begin{equation} \label{eq:eff_eta4}
    \eta_{\rm eff} = \frac{1}{J_0}\sqrt{E_0^2 + e_z^2} \; ,
\end{equation}
where 
\begin{equation}
\vec{e} =  \Gamma \left[\,\vec{E} + \frac{1}{c}\vec{v}\times\vec{B}\right]
\end{equation}
is the electric field in the fluid frame \citep[e.g.][]{mignone2019}.

Compared to \refeq{eq:eff_eta3}, this expression for the effective resistivity has no singularities and is completely determined by fluid quantities readily available at each time step (whereas the current density would in principle require to know the instantaneous value of the displacement current).
More importantly, \refeq{eq:eff_eta4} has the benefit of producing a profile of resistivity that systematically peaks within the current sheet and tends to smoothly decrease once dissipation uniforms the magnetic field profile (see Appendix~\ref{app:komissarov} for a 1D numerical test), thus leading to a physically motivated dynamics despite the strong approximations used in its derivation.

%-------------------------------------------------------------------------------

\subsubsection{Numerical implementation}\label{subsec:eta_implementation}
\refeq{eq:eff_eta4} needs to be properly expressed in a dimensionless form, if one wants to employ it within a resistive RMHD code.
This is due to the fact that the fluid equations are scale-free in the limit of a constant scalar resistivity that does not depend on fundamental physical constants.
We indicate with $L_u$ and $\rho_u$ respectively the unit length and mass density, which we use to rescale the variables appearing in \refeq{eq:eff_eta4} in terms of dimensionless quantities (indicated in the following with a bar notation)
\begin{equation}
    \eta_{\rm eff} = \frac{4\pi}{\omega_u \bar{\rho}} \sqrt{\left(\bar{\delta}_u\partial_{\bar{y}} \bar{v}_y \right)^2 + \bar{e}_z^2} \; .
\end{equation}
In the previous expression we rescaled the electric (and magnetic) field according to $\vec{E} = \sqrt{4\pi \rho_u c^2} \vec{\bar{E}}$ and we introduced the dimensionless plasma skin-depth $\bar{\delta}_u=c/(\omega_uL_u)$, with $\omega_u=\sqrt{\frac{4\pi e^2 \rho_u}{m^2}}$ being the plasma frequency associated to the chosen unit of mass density.
We can then obtain an expression for the dimensionless resistivity $\bar{\eta}_{\rm eff}$ by factoring out the physical units in Amp\`ere-Maxwell's law
\begin{equation}
    \frac{\partial \bar{\vec{E}}}{\partial \bar t} - \bar{\vec{\nabla}} \times \bar{\vec{B}} = - \bar{\vec{J}} \equiv - \bar{\rho}_e \bar{\vec{v}} - \frac {1} {\bar{\eta}_\mathrm{eff}} \Gamma \left[ \bar{\vec{E}} + \bar{\vec{v}} \times \bar{\vec{B}} - (\bar{\vec{E}} \cdot \bar{\vec{v}}) \bar{\vec{v}} \right] \; ,
\end{equation}
which leads to this final form for the effective resistivity employed in the \pluto{} code
\begin{equation}\label{eq:eff_eta5}
    \bar{\eta}_{\rm eff} \equiv \frac{c}{4\pi L_u}\eta_{\rm eff} =   \frac{\bar{\delta}_u}{\bar{\rho}} \sqrt{\left(\bar{\delta}_u\partial_{\bar{y}} \bar{v}_y \right)^2 + \bar{e}_z^2}\; .
\end{equation}
From \refeq{eq:eff_eta5} we can see how adopting the effective resistivity prescription introduces a characteristic scale to the problem, $\bar{\delta}_u$, which is set by the specific choice of physical units for lengths and mass densities adopted by the code.
This length scale represents a crucial parameter that determines the magnitude of the effective resistivity and has a clearly defined physical meaning.
For this reason, when we use \refeq{eq:eff_eta5}, we set the amount of magnetic dissipation not by arbitrarily choosing a constant resistivity $\eta_0$, but rather properly setting an equivalent plasma skin-depth length scale and a density profile for the current sheet.
Since the value of $\bar{\delta}_u$ is a direct consequence of the physical units adopted by the code and the initial scale of the problem, we treat it as a fixed parameter that does not dynamically change with the local plasma skin depth (the dependence on $\rho$ of $\eta_{\rm eff}$ already captures this effect).
We compute the spatial derivative of the velocity component transverse to the current sheet via central differences and applying a minmod limiter to avoid spurious oscillations that would otherwise affect the resistivity profile. 
We also set a floor value of $\eta_{\rm floor}=10^{-6}$ in order to avoid excessively small values of resistivity that would simply hinder the stability of the code without significantly affecting the simulation's dynamics \citep{ripperda2019b}. 

%-------------------------------------------------------------------------------

\subsection {PIC models}
Our simulations of collisionless magnetic reconnection are produced with the 2D version of the relativistic electromagnetic PIC code \texttt{Zeltron} \citep{cerutti2013,cerutti2019}, which solves the time-dependent Maxwell equations
\begin{eqnarray}
    \frac{\partial\vec{E}}{\partial t} & = &c{\vec{\nabla}}\times \vec{B} - 4\pi\vec{J}  \\
    \frac{\partial\vec{B}}{\partial t} & = &-c{\vec{\nabla}}\times \vec{E}\; ,  
\end{eqnarray}
using a second-order finite-difference time-domain method \citep{yee1966}.
While the divergence-free condition for the magnetic field is ensured to machine precision at any given time, the Gauss equation $\vec{\nabla}\cdot\vec{E}=4\pi\rho_e$ is enforced by applying at each time step a correction $\delta \vec{E}$ to the electric field obtained by solving Poisson's equation
\begin{equation}
    \nabla^2(\delta\phi) = -(4\pi\rho_e-\vec{\nabla}\cdot\vec{E})\; ,
\end{equation}
where $\rho_e$ is the electric charge density and $\delta \vec{E}=-(\vec{\nabla}\delta\phi)$.
The electromagnetic field is coupled to the dynamics of the charged particles through their equation of motion
\begin{equation}
    m\frac{d\vec{u}_p}{dt} = q\left(\vec{E}_p+\frac{\vec{u}_p\times\vec{B}_p}{c\Gamma_p}\right)\; ,
\end{equation}
where $m$ is the particle's mass, $\vec{u}_p=\Gamma_p\vec{v}_p$ is its 4-velocity, $\vec{v}_p$ is its spatial velocity, $q$ is its electric charge, $\Gamma_p = \sqrt{1 + (u_p/c)^2}$ is its Lorentz factor, and $\{\vec{E}_p,\vec{B}_p\}$ are the electric and magnetic fields at the particle's location.
The equation of motion is integrated using the classic Boris pusher \citep{birdsall1991}. Charges and currents generated by each particle are then deposited on the nodes of the grid via linear interpolation, where they are used to compute the charge and current densities for Maxwell's equations.
In this work, we will assume a pure electron-positron pair plasma as in \citet{selvi2023}.

%-------------------------------------------------------------------------------

%
\begin{figure*}
    \centering
    \includegraphics[width=\textwidth]{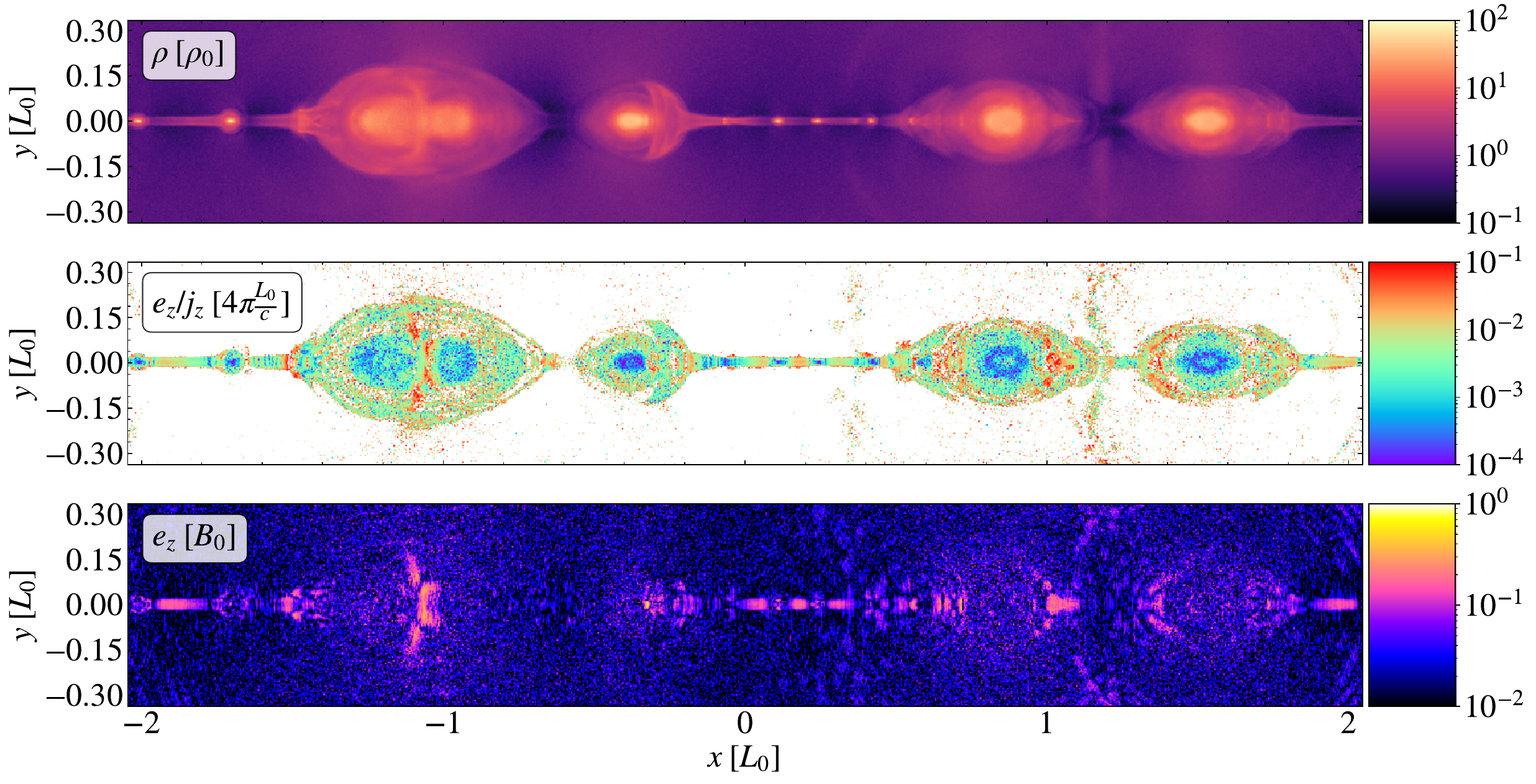}
    \caption{
    Spatial profiles of (from top to bottom) mass density, ratio of comoving electric field over current density $e_z/j_z$, and $e_z$ for the kinetic model \pic{10} at $t \simeq 2.4$.
    Density is normalized by the upstream value $\rho_0$, the electric field by the upstream magnetic field $B_0$, and the current density by $J_0=\frac{c}{4\pi}\frac{B_0}{L_0}$.
    }
    \label{fig:2d_snapshots_PIC}
\end{figure*}
\begin{figure}
    \centering
    \includegraphics[width=0.48\textwidth]{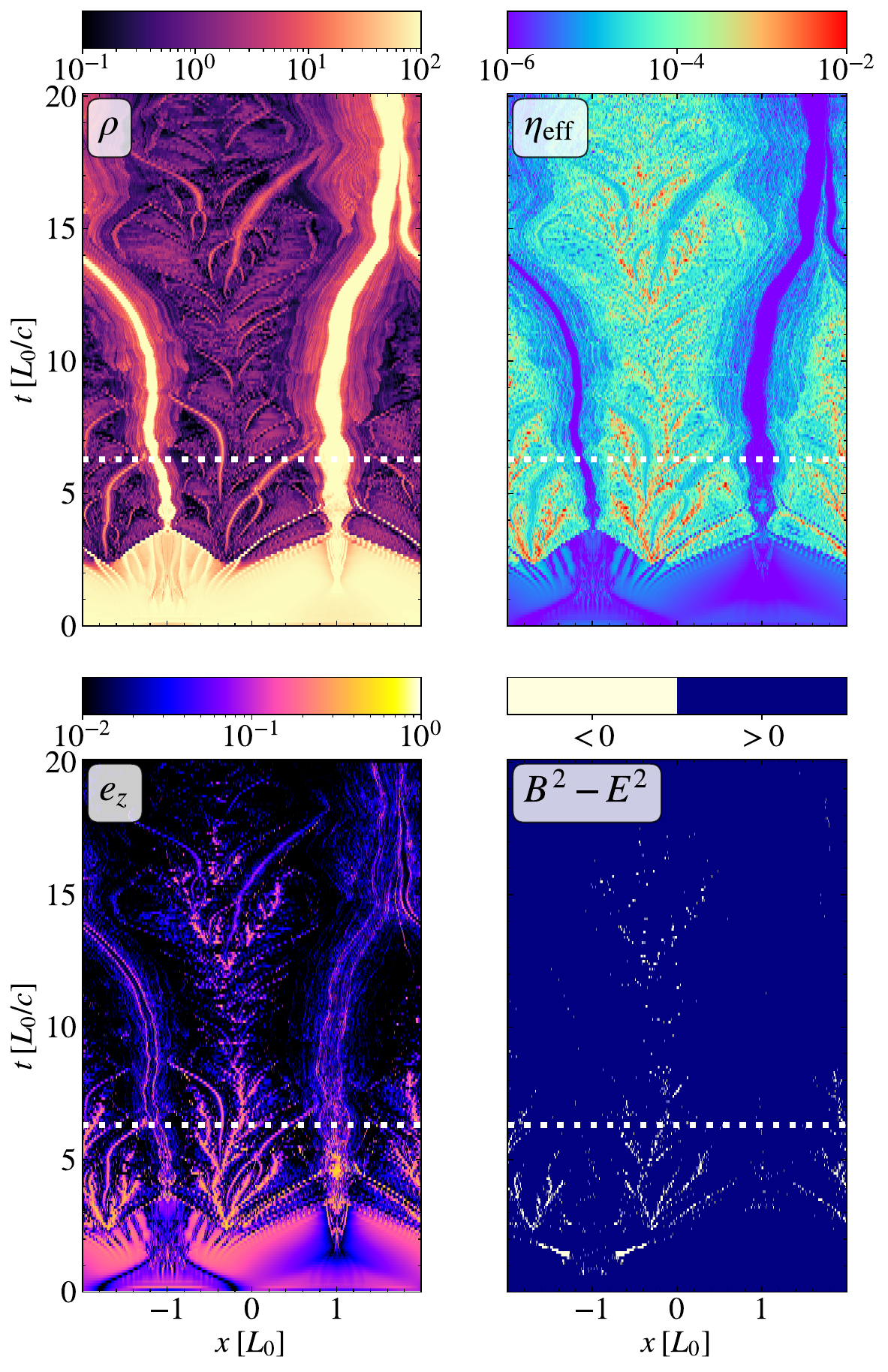}
    \caption{Space-time diagrams of different quantities at the center of the current sheet ($y=0$) over time for the ResRMHD model \rmhd{10}{E}{1}. 
    From top left to bottom right: rest-mass density $\rho$, effective resistivity $\eta_{\rm eff}$, $z-$component of the electric field in the fluid frame $e_z$, and Lorentz invariant $B^2-E^2$.
    The white dotted lines indicate the instant at which the profiles in \refig{fig:2d_snapshots_lowdelta} are shown.}
    \label{fig:spacetime_diagrams_lowdelta}
\end{figure}

\subsection{Initial conditions}
We initialize our domain with a standard Harris current sheet of width $a$. 
In the following, quantities with subscript ``$_0$'' refer to the upstream region of the domain, i.e. far from the current sheet.
The magnetic field has non-vanishing component only along the $x-$axis, which is defined as
\begin{equation}
    B_x(y) = B_0 \tanh\left(\frac y a\right)\, ,
\end{equation}
where $B_0$ is set by the plasma ``cold'' magnetization to $\sigma_0=B_0^2/\rho_0$ and the plasma density $\rho_0$
(here $\vec{B}/\sqrt{4\pi}\to\vec{B}$ for convenience).
Once we choose the ratio of thermal to magnetic pressure $\beta_0=2p_0/B_0^2$ (where $p_0$ is the upstream plasma thermal pressure), we can compute the Alfv\'en speed as in \cite{delzanna2016}
\begin{equation}
    c_{A,0} = (1/\sigma_0+2\beta_0+1)^{-1/2}\, ,
\end{equation}
where we assumed an ideal equation of state for a relativistic plasma with adiabatic index 4/3.
The thermal pressure $p$ is then computed by requiring a constant total pressure $p_{\rm tot}=p+B^2_x/2$ in all the domain,  thus giving
\begin{equation}
    p(y) = \frac 1 2 B_0^2 (\beta + 1) - \frac 1 2 B_x^2(y)\, ,
\end{equation}
which we then use to compute the density across the domain by assuming an initial constant temperature $\Theta = p/\rho = 1$.
This choice leads the upstream plasma density $\rho_0$ to increase by a factor $p(0)/p_0=(\beta+1)/\beta\simeq\beta^{-1}=100$ within the current sheet. 

The magnetic field is expressed in terms of a vector potential along the $z$ direction 
\begin{equation}\label{eq:az_init}
    A_z(x,y) = aB_0 \log(\cosh{y/a}) + \frac {\epsilon B_0L_0} {2\pi} \sin \left(\frac{2\pi x}{L_0}\right) \text{sech}\left(\frac{y}{a}\right)
\end{equation}
where the second term of the $z-$component introduces a sinusoidal perturbation with wavelength $L_0$ and amplitude $\epsilon=0.05$ that is localized in proximity of the current sheet to seed the tearing instability in the ResRMHD models and predefine the location of the main $X$ and $O$ points along the sheet.

Our numerical domain consists of a Cartesian box with a uniform grid spacing equal to $\Delta x\simeq\delta_0/20$ (unless otherwise stated), where $\delta_0=c/\omega_0$ is the upstream plasma skin-depth and $\omega_0$ the corresponding plasma frequency.
Our box covers the range $[0,4L_0]$ in the $x-$direction, where $L_0=1$ is the characteristic length of the current sheet, while we set the current sheet width $a=0.01L_0=\delta_0/2$.
This length is 500 times the current sheet skin-depth $\delta_p$, since $\omega_p \simeq 10 \omega_0$.
Throughout this work we consider values for the dimensionless skin-depth parameter $\bar{\delta}_u$ in \refeq{eq:eff_eta5} corresponding to either the upstream or the current sheet plasma skin-depth, i.e. $\bar{\delta}_u=\{\delta_0/L_0,\delta_p/L_0\}=\{0.02,0.002\}$.
Imposing a specific value to $\bar{\delta}_u$ and assuming $L_0=L_u$ is equivalent to setting the unit of mass density to $\rho_u = m/r_e (\bar{\delta}_uL_u)^{-2}$, with $r_e$ the classical electron radius.
Since the rest of the MHD equations remain scale-free, this adjustment does not affect any further the dynamics of the system.
The size of our numerical box in the direction transverse to the current sheet needs to be sufficiently large so as to avoid artificial high-energy cut-offs in the spectra of particles accelerated by the reconnection electric field.
\citep{werner2016} showed that when the size of the domain in the direction transverse to the current sheet is not sufficiently large, i.e. when
\begin{equation}\label{eq:werner2016}
L_y\lesssim40\sigma_0 r_L\, , 
\end{equation}
the spectra of particles with mean Larmor radius $r_L$ that are accelerated by the reconnection electric field exhibit an artificial high-energy cut-off.
We conservatively double the constrain from \refeq{eq:werner2016} and set $y\in[-80\sqrt{\sigma_0}a,80\sqrt{\sigma_0}a]$, where we used the fact that $\delta=\sqrt{\sigma_0}r_L$.  
We impose in the $x-$ and $y-$direction periodic and reflective boundary conditions, respectively. 

In the PIC simulations, the plasma is generated with 32 particles per cell everywhere in the domain. 
The initial density gradient in the layer is achieved thanks to variable particle weights. 
The initial particle energy distribution follows a relativistic Maxwellian at rest in the upstream medium, while the plasma in the layer is generated according to two counter-streaming relativistic drifting Maxwellian carrying the current (one beam for each species drifting at $v_{\rm drift}=\pm 0.6 c$ along the $z$-direction).
All kinetic models start with the same initial plasma skin depth as the ResRMHD case (i.e. $0.02L_0$ upstream, $0.002L_0$ at the center of the current sheet), the same numerical box, and a uniform grid spacing of $\Delta x = \delta_0/10$.

\begin{figure*}
    \centering
    \includegraphics[width=\textwidth,clip=True]{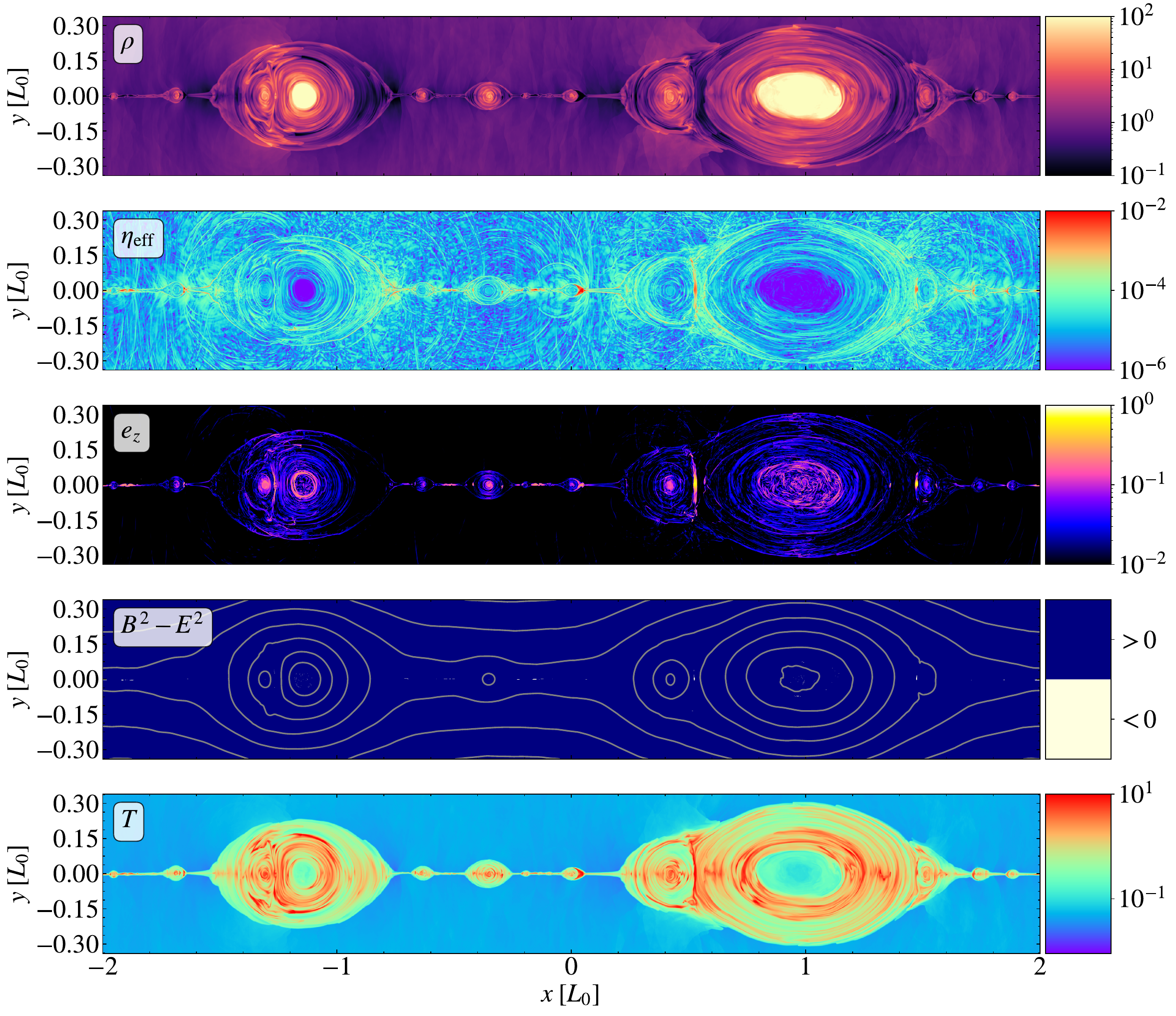}
    \caption{
    Spatial profiles of (from top to bottom) the rest-mass density $\rho$, the effective resistivity $\eta_{\rm eff}$, the $z-$component of the electric field in the fluid frame $e_z$, the Lorentz invariant $B^2 - E^2$ with the contours of the vector potential $A_z$, and the temperature for the ResRMHD model \rmhd{10}{E}{1} at $t=6.3$.
    }
    \label{fig:2d_snapshots_lowdelta}
\end{figure*}
\begin{figure}                              
    \centering
    \includegraphics[width=0.48\textwidth]{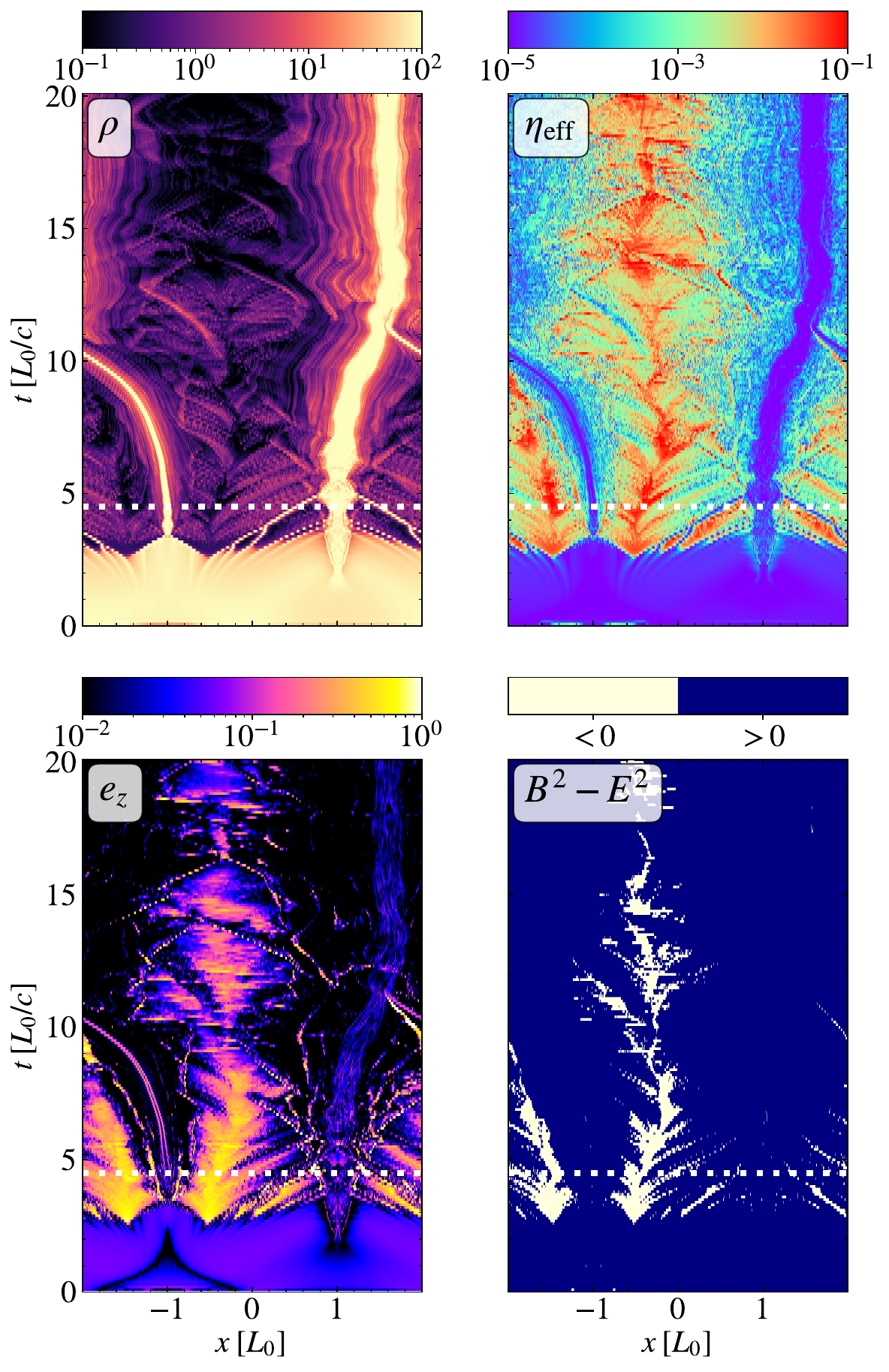}
    \caption{Same as \refig{fig:spacetime_diagrams_lowdelta}, but for the ResRMHD model \rmhd{10}{Ed0}{1}.
    The white dotted lines indicate the instant at which the profiles in \refig{fig:2d_snapshots} are shown.
    }
    \label{fig:spacetime_diagrams_highdelta}
\end{figure}
\begin{figure*}
    \centering
    \includegraphics[width=\textwidth]{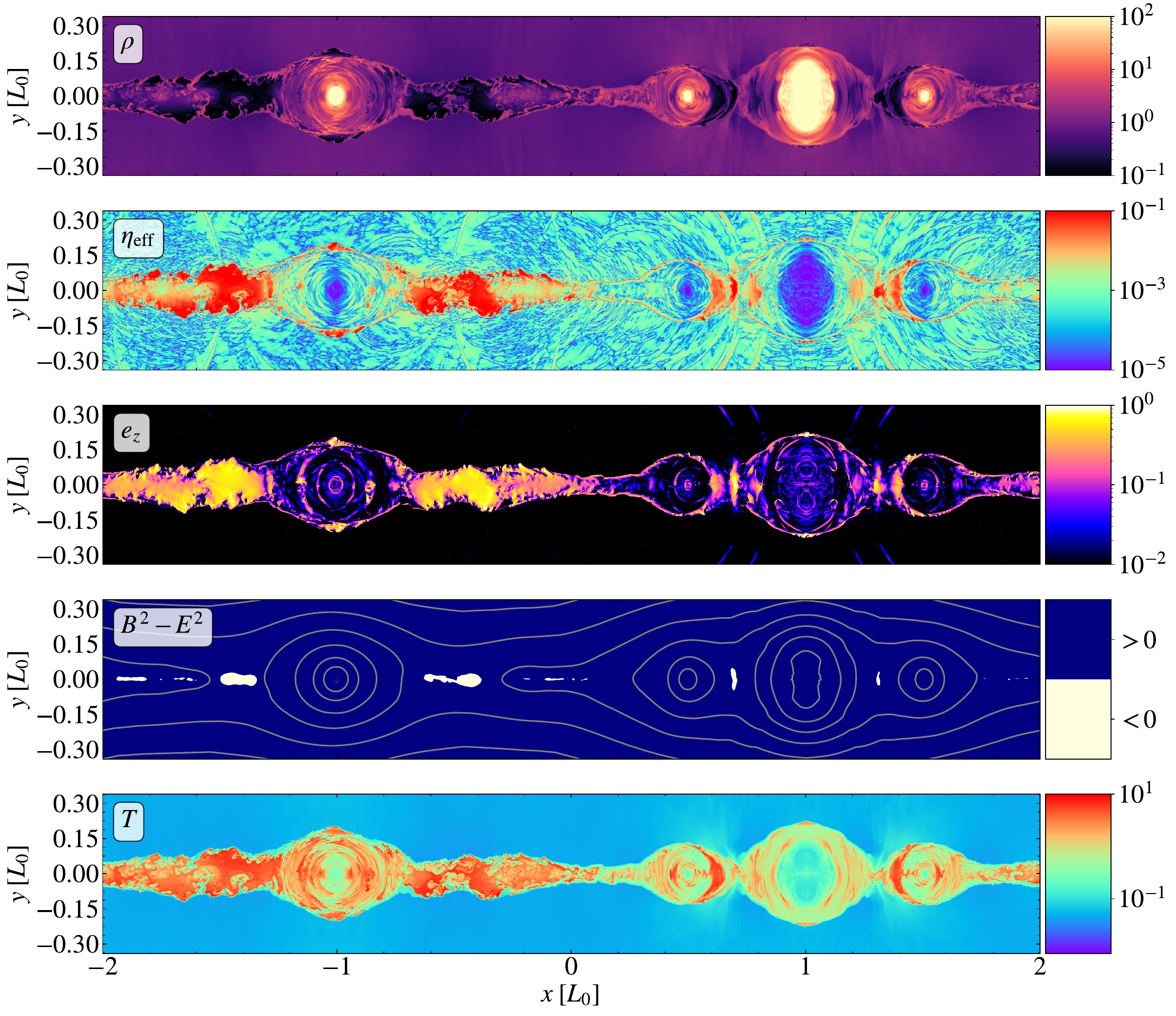}
    \caption{
    Same as \refig{fig:2d_snapshots_lowdelta}, but for the ResRMHD model \rmhd{10}{Ed0}{1} at $t=4.5$.
    }
    \label{fig:2d_snapshots}
\end{figure*}
%
% %%%%%%%%%%%%%%%%%%%%%%%%%%%%%%%%%%%%%%%%%%%%%%%%%%%%%%%%%%%%%%%%%%%%%%%%%%%%%%
%%%%%%%%%%%%%%%%%%%%%%%%%%%%%%%%%%%%%%%%%%%%%%%%%%%%%%%%%%%%%%%%%%%%%%%%%%%%%%%%
%%%%%%%%%%%%%%%%%%%%%%%%%%%%%%%%%%%%%%%%%%%%%%%%%%%%%%%%%%%%%%%%%%%%%%%%%%%%%%%%
%%%%%%%%%%%%%%%%%%%%%%%%%%%%%%%%%%%%%%%%%%%%%%%%%%%%%%%%%%%%%%%%%%%%%%%%%%%%%%%%
%%%%%%%%%%%%%%%%%%%%%%%%%%%%%%%%%%%%%%%%%%%%%%%%%%%%%%%%%%%%%%%%%%%%%%%%%%%%%%%%

\section{Effective resistivity models}\label{sec:analysis}
\subsection{Current sheet dynamics}\label{sec:general_analysis}
We first briefly describe the main properties of kinetic model \pic{10}, which represents a realization from first principles of the reconnection of the current sheet we consider in this study.
In the following we measure space and time in units of $L_0$ and $L_0/c$, respectively. 
The initial equilibrium is promptly destroyed by a violent fragmentation of the current sheet in multiple magnetic islands that start merging and producing ever larger plasmoids, with a single large magnetic island remaining once saturation is reached.
\refig{fig:2d_snapshots_PIC} shows a snapshot taken at $t\simeq2.4$ that showcases this well-known dynamics.
A chain of four large high-density magnetic islands is interrupted by a series of small plasmoids that are about to merge with the larger $O-$points (top panel).
Regions of high resistive electric field develop in between plasmoids and at the interface of merging magnetic islands (bottom panel).
The ratio between such electric field component and the current density (middle panel) represents the quantity that we refer to as the effective resistivity of the system.
To avoid spurious high values in the upstream regions due to exceedingly low current densities, we do not render this ratio wherever the current density's magnitude drops below a factor $10^{-2}$ with respect to its maximum at the beginning of the simulation.
We can see a clear anti-correlation with mass density in the low resistivity regions at the core of each plasmoid, while high values of $e_z/j_z$ are reached in correspondence to high values of the resistive electric field.
This relations are both in agreement with the dependencies included in \refeq{eq:eff_eta5} we previously introduced.  

We now analyze models \rmhd{10}{E}{1} and \rmhd{10}{Ed0}{1} to provide a reference for the reconnection dynamics produced by the effective resistivity prescription (a complete list of the simulations included in this work is presented in Table~\ref{tab:models}).
Here we consider a simulation with $N_x=4096$ points along the $x-$axis (which corresponds to $\Delta x=\delta_0/20$, thus with twice the resolution used in the corresponding PIC simulation), leaving a deeper analysis of the impact of numerical dissipation for Section \ref{subsec:res}.  
As a reference, for a given resolution the PIC models require about twice as much computational resources with respect to the ResRMHD models.
The employment of AMR in the grid structure would have further reduced the computational cost of the fluid models (especially in the upstream region, where it is enough to use a coarser resolution), but we opted for a uniform grid in order to more closely match the one used by PIC simulations.

% Spacetime diagrams low delta
In \refig{fig:spacetime_diagrams_lowdelta} we show a space-time diagrams of several quantities at the center of the reconnecting layer \citep[similarly to Fig.~2 of ][althought for ResRMHD simulations rather than PIC]{nalewajko2015}.
At the beginning of the simulation the current sheet becomes thinner by $\simeq20\%$ due to the onset of the tearing instability, with the initial perturbation inducing the formation of a small magnetic island at $x=1$ around $t\simeq1.7$ (top left panel of \refig{fig:spacetime_diagrams_lowdelta}).
During this phase, the resistivity remains relatively low ($\sim10^{-5}$) close to the two large $X$ and $O$ points forming in $x=\pm1$ (top right panel). 
After this phase the tearing instability enters its non-linear stage, with the current sheet beginning to also fragment into smaller plasmoids.
The resistivity increases by more than an order of magnitude in between the smaller magnetic islands (where density decreases), as does the electric field in the plasma comoving frame.
At $t\simeq2.5$ the initial chain of small plasmoids starts to merge towards $x=\pm1$, leaving by $t\simeq4$ two large magnetic islands towards which newly formed plasmoids continue to quickly merge throughout the simulation.
Both the resistivity and the non-ideal electric field spike in the low-density regions of the current sheet's center at the sides of the plasmoids, with a clear correlation between them (top right and bottom left panels of \refig{fig:spacetime_diagrams_lowdelta}).
These regions, which are sparsly distributed, produce also favorable conditions for the acceleration of particles, as indicated by the negative values assumed by the Lorentz invariant $B^2-E^2$ \citep{sironi2022}.
The $O-$point formed in $x=-1$ around $t\simeq4$ approaches instead slowly the lower periodic boundary of the $x$ axis and merges across it with the original island around $t\simeq18$.
This halts the further reconnection of magnetic field lines, which leads to a drop of the non-ideal electric field on the sides of the last plasmoid and the relaxation towards a global saturated state with a single large plasmoid.

% 2D snapshots low delta
\refig{fig:2d_snapshots_lowdelta} shows the profiles of a series of quantities at $t=6.3$ for model \rmhd{10}{E}{1}, when the two largest plasmoids are steadily merging with smaller ones.
The morphology of the reconnecting layer is consistent with the standard scenario produced by both kinetic and fluid models: the fragmentation of the current sheet produces a continuous coalescence of newly formed small structures into larger magnetic islands, with thin current filaments connecting them.
The effective resistivity is larger around and in between the plasmoids, with strong peaks emerging at the interface of merging islands and in agreement with the results obtained by model \pic{10}.
The non-ideal electric field also reaches its maximum value in these regions, but it is otherwise stronger within the magnetic islands, rather than at the edges.
These regions also host the most promising sites for the initial phases of particle acceleration, with negative values of $B^2-E^2$ reached in small patches in the middle of highly resistive parts of the current sheet. 
The last panel of \refig{fig:2d_snapshots_lowdelta} shows how the highest temperatures are reached not only around the dense core of the two largest plasmoids, but also in the rarefied regions of strong magnetic dissipation and non-ideal electric field, followed by the regions between magnetic islands where the electric field and the resistivity both peak.  

% Spacetime diagrams high delta
Model \rmhd{10}{Ed0}{1}, for which $\bar{\delta}_u$ is ten times larger, shows a similar initial transient in terms of current sheet thinning and fragmentation.
However, both resistivity and electric field are initially weaker than the previous case (top right and bottom left panels of \refig{fig:spacetime_diagrams_highdelta}).
In the non-linear stage of the simulation we still have two large plasmoids forming at $x=\pm1$ and merging together on the right side of the numerical box, but the process is faster and leads, by $t\simeq 12L_0/c$, to a saturated state.
Both resistivity and electric field correlate still very well, but are now much stronger and cover larger fractions of the current sheet's length.
This is accompanied by a clear lack of production of short-lived small plasmoids between the two main magnetic islands, where density generally drops by more than an order of magnitude.  
The regions with favorable conditions for particle acceleration (bottom right panel) are now much larger and last longer in time, consistently with the aforementioned dissipation dynamics.

% 2D snapshots high delta
The spatial profiles of such quantities at $t=4.5$ (when the activity within the current sheet is most intense) are also qualitatively different than the case using $\bar{\delta}=0.002$.  (\refig{fig:2d_snapshots}).
Both resistivity and non-ideal electric field reach their maximum values in the regions of low density (i.e. between the high-density merging plasmoids), which is consistent with the effective resistivity being inversely proportional to $\rho$.
However, rather than having thin patches in between merging magnetic islands, we now have large, rarefied cavities with turbulent edges, which disrupt the current sheet and prevent the formation of chains of small-scale plasmoids. 
These regions also host the most promising sites for particle acceleration, with negative values of $B^2-E^2$ reached in patches that are considerably larger than those produced in model \rmhd{10}{E}{1}. 
The large temperatures reached between plasmoids (last panel) confirm that such bubbles are produced by an excess of magnetic dissipation due to the much larger resistivity that comes with a larger value of $\bar{\delta}$.
Such evolution deviates significantly from the reconnection dynamics produced by the self-consistent PIC model, hence representing a limiting case for the effective resistivity formulation.
This effect suggests that the layer thickness is set by the plasma skin depth scale in the sheet rather than in the upstream medium, which makes more sense physically.
Model \rmhd{10}{E}{1} will therefore serve as our reference run for the rest of the paper.

% RMS By
\subsection{Global evolution of the magnetic field}
\begin{figure}[h!]
    \centering
    \includegraphics[width=.48\textwidth]{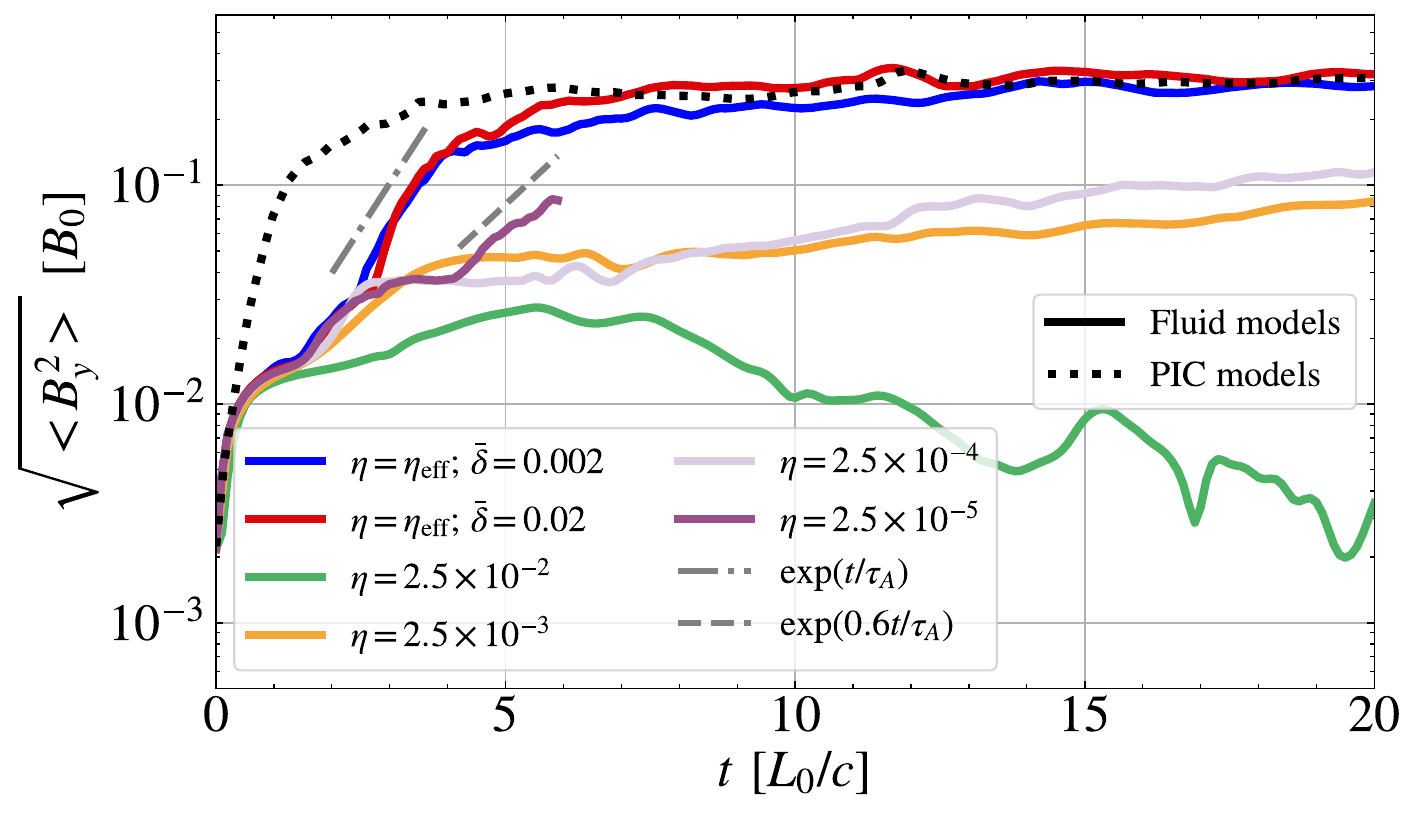}
    \includegraphics[width=.48\textwidth]{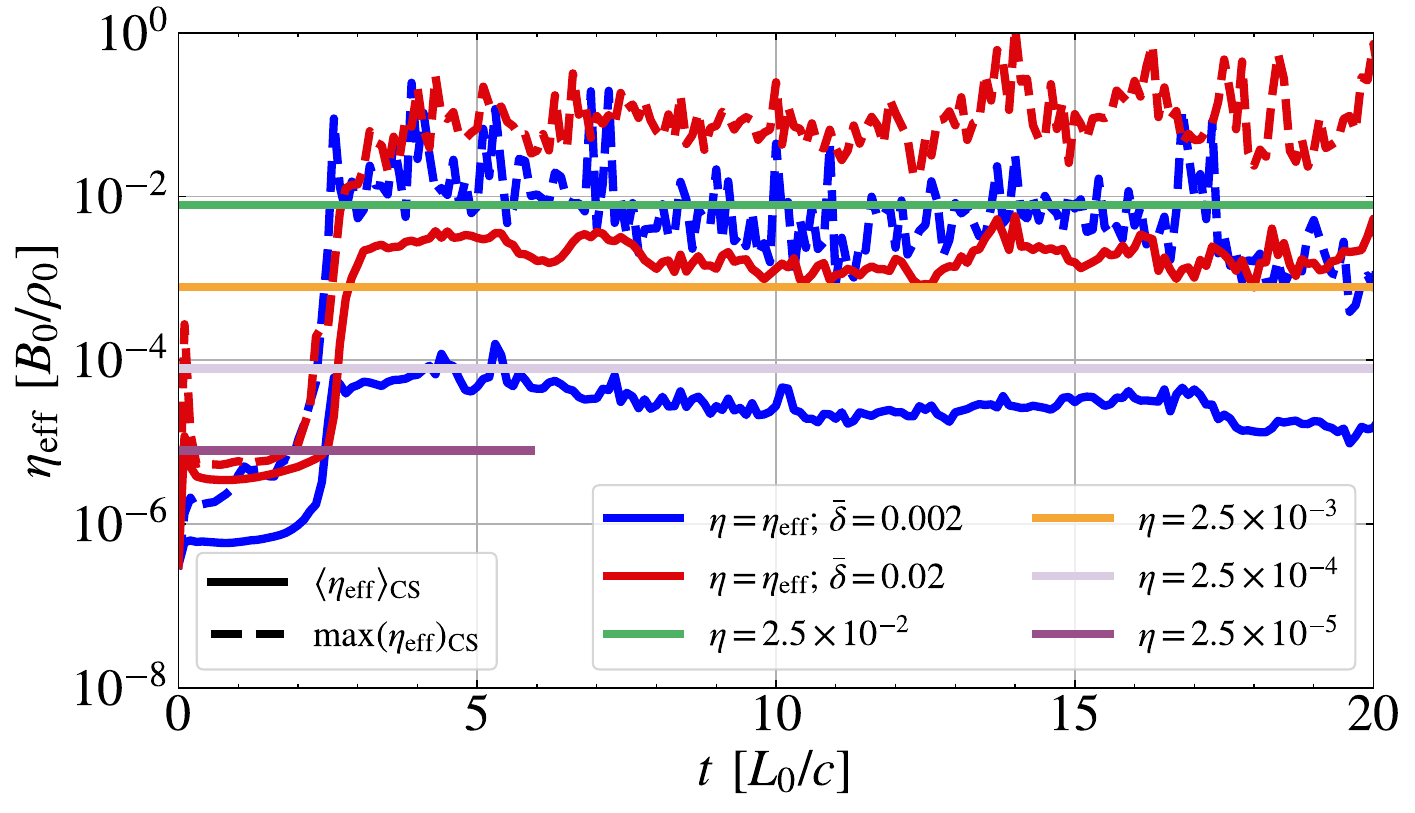}
    \caption{Top: RMS value of the transverse component of the magnetic field over time for ResRMHD models \rmhd{10}{E}{1} (blue curve), \rmhd{10}{Ed0}{1} (red), models with constant resistivity (green, yellow, light purple, and purple), and model \pic{10} (dotted).
    Bottom: average and maximum value of the resistivity over time.
    The average is calculated over a rectangular box of sizes $L_x\times 2a$ centered in $y=0$.}
    \label{fig:by_eta_eta}  
\end{figure}

The onset of the tearing instability triggered by the magnetic seed perturbation leads to an initial fast growth of the RMS $y-$component of the magnetic field (top panel of \refig{fig:by_eta_eta}).
The initial transient corresponding to the linear phase of the instability lasts roughly until $t\simeq2.5$, at which point the fragmentation of the current sheet into smaller plasmoids induces an even faster increase of the reconnected magnetic field.
The growth rate in this stage is roughly the inverse of the ideal time scale $\tau_A=L_0/c_A$, which is even faster than the value predicted by the ``ideal tearing'' regime \citep{pucci2014,landi2015}, i.e. $\simeq 0.6 \tau_A^{-1}$.
Such value corresponds, in principle, to a specific choice of aspect ratio for the current sheet (i.e. $L/a=S^{1/3}$) and becomes smaller for thicker current sheets according to \citep{delzanna2016,komissarov2024}
\begin{equation*}
    \gamma_{\rm max}\tau_A \simeq 0.6 S^{-1/2} (a/L)^{-3/2}\; .
\end{equation*}
However, since the system is unstable on dynamical time scales when the current sheet reaches this thickness, obtaining a smaller value of a/L (hence faster reconnection) is hindered by the fragmentation of the current sheet itself.
The growth rate $\gamma_{\rm max}=0.6\tau_A^{-1}$ can therefore be considered an upper limit to what can be produced by a reconnection event in the ResRMHD regime with constant resistivity.
After $t\simeq7.5$ there is no significant growth of this quantity for model \rmhd{10}{Ed0}{1} as the system reaches a fully saturated state.
The model with a lower value for $\bar{\delta}$ produces almost the same evolution for $\langle B_y^2\rangle$, with only a marginally slower growth after $t\simeq4$.
Overall, the corresponding PIC simulation (black dotted line) produces a very similar evolution as well.   
The growth of $\langle B_y^2 \rangle$ in the kinetic model is fast and steady, with a rate of increase close to the one experienced by the ResRMHD simulations after $t\simeq2.5$, i.e. the moment when their current sheet starts fragmenting.
When the RMS field reaches the value $\sim0.1B_0$, the field in the PIC simulation quickly saturates to the same intensity as for the fluid models ($\simeq0.3B_0$).
The difference between the initial transients occurring in the fluid and kinetic models stems from the fact that the former needs to go through a thinning of the current via the tearing instability, before reaching the point where the first plasmoids can form.
On the other hand, the current sheet in model \pic{10} fragments immediately, therefore producing without delays a powerful reconnection event.

%Resistivity over time
As previously seen, the effective resistivity assumes a wide range of values during the simulation.
In the bottom panel of \refig{fig:by_eta_eta} we show the evolution over time of the average resistivity in the current sheet (specifically, within a distance $a$ from the center) and the maximum value it assumes within the same region for models \rmhd{10}{E}{1} and \rmhd{10}{Ed0}{1}.
We report it in units of $B_0/\rho_0$, as is it a quantity appearing from \refeq{eq:eff_eta5} that provides a natural scaling unit.
During the linear phase of the tearing instability the average resistivity increases immediately from the initial floor value of $\simeq3\times10^{-7}B_0/\rho_0$ due to the local resistive electric field building up (bottom left panels of \refig{fig:spacetime_diagrams_lowdelta} and \refig{fig:spacetime_diagrams_highdelta}).
The increase ranges from a few times (model \rmhd{10}{E}{1}) and 10 times (model \rmhd{10}{Ed0}{1}), depending on the value of $\bar{\delta}$.
Around $t\simeq2.5$ the resistivity then grows by up to three orders of magnitude due to the current sheet beginning to fragment, with a steady mean value at the center of the current sheet of roughly either $3\times10^{-5}B_0/\rho_0$ (\rmhd{10}{E}{1}) or $2\times10^{-3}B_0/\rho_0$ (\rmhd{10}{Ed0}{1}), with a difference between the two models larger than the ratio between the two plasma skin-depths used by these models.
This is due to the fact that the non-ideal electric field in model \rmhd{10}{Ed0}{1} is stronger than the one produced with a smaller value of $\bar{\delta}$, hence increasing more the local dissipation rate.
After this fast increase in resistivity, both average and maximum values do not vary much with time, with a ratio between the two up to $\sim500$.
However, in model \rmhd{10}{E}{1} the maximum value of $\eta_\mathrm{eff}$ right after the formation of the first plasmoid chains ($t\simeq2.5$) decreases steadily by a an order of magnitude over the period required by the reconnected magnetic field to saturate, whereas such decrease does not occur in the case of larger $\bar{\delta}$.

% Reconnected magnetic flux
To quantify the efficiency of reconnection in our models, we first compute for both ResRMHD and PIC simulations the reconnected magnetic flux $\Phi_{\rm rec}$ along the current sheet ($y=0$) as
\begin{equation}
    \Phi_{\rm rec} = - \int_{x_X}^{x_O} B_y(x,0)dx  = A_z\big\vert_X - A_z\big\vert_O\; ,
\end{equation}
where we identify the position of the $X$ ($O$) point by the maximum (minimum) value of $A_z$.
For both models with effective resistivity, the reconnected magnetic flux starts increasing significantly from $t\simeq2.5$, i.e. when the current sheets start fragmenting.
Model \rmhd{10}{E}{1} produces a rather steady growth of the magnetic flux, slowly saturating towards $\Phi_{\rm rec}\simeq0.5\Phi_0$ by $t\simeq40$, where $\Phi_0=B_0L_y/2$ is a measure of the magnetic flux available for reconnection at the beginning of the simulation.
The flux from model \rmhd{10}{Ed0}{1} increases instead at a faster pace, it slows down after $t\simeq8.5$ (due to the merger between the two last plasmoids being slower at the beginning) and then saturates at a similar value.
Such deviation from the model with lower $\bar{\delta}$ reflects the faster and more violent reconnection dynamics produced by this model.
Similarly to the evolution of $B_y$, the PIC simulation shows an immediate growth of the reconnected flux shortly after the beginning of the simulation, since the current sheet does not undergo a preliminary linear stage of the tearing instability.
Besides this small offset, however, the evolution of the reconnected flux of the PIC model is very similar to that of the ResRMHD ones.
The initial growth of $\Phi_{\rm rec}$ falls roughly in between the two fluid models, halting briefly before the merger of the two last large magnetic islands and then saturating close to $0.5\Phi_0$ as well.

We also compute the corresponding reconnection rates $\beta_{\rm rec}$ as
\begin{equation}
    \beta_{\rm rec} = \frac{v_{\rm rec}}{c_{A,0}} = \frac{1}{B_0c_{A,0}}\frac{d\Phi_{\rm rec}}{dt}\; ,
\end{equation}
where we identified the inflow velocity of the magnetic field lines into the $X$-point with $v_{\rm rec}=B_0^{-1}d\Phi_{\rm rec}/dt$.
Besides the delay between PIC and ResRMHD models, the reconnection rates obtained with the effective resistivity formulation are remarkably similar to the kinetic result.
In the \rmhd{10}{Ed0}{1} simulation $\beta_{\rm rec}$ spikes to a maximum value of $\simeq0.2$ at the start of the current sheet fragmentation that lasts until $t\simeq6$.
Model \rmhd{10}{E}{1} produces a qualitatively comparable scenario, with a peak around $\simeq0.1$ and a steadier mean value over time, which is consistent with the less dramatic increase in resistivity with respect to the case having $\bar{\delta}=0.02$.
Finally, PIC and ResRMHD models exhibit a very similar time variability, on both shorter and longer time scales. 

\begin{figure}
    \centering
    \includegraphics[width=.48\textwidth]{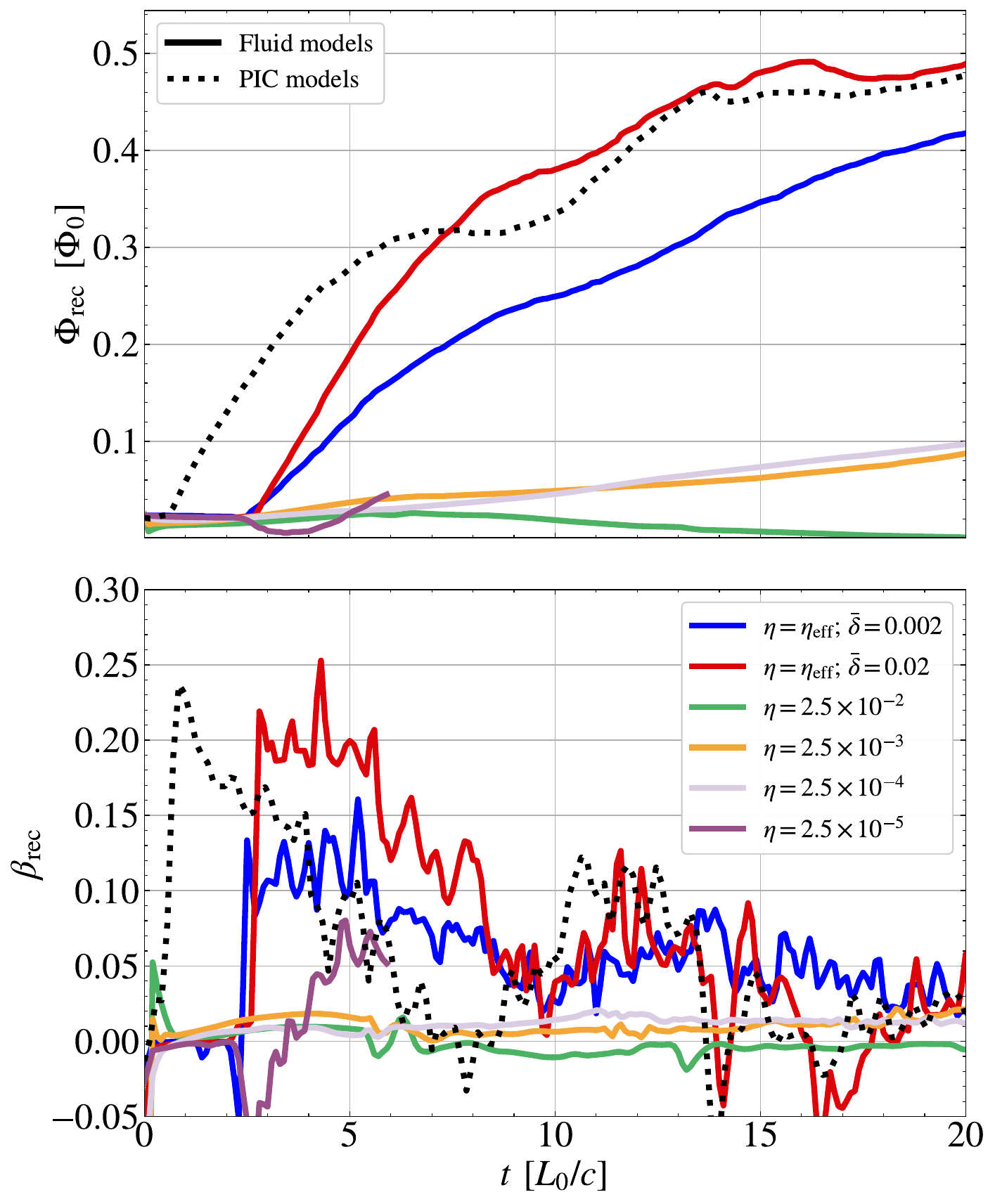}
    \caption{Reconnected magnetic flux (top) and reconnection rate (bottom) over time for the same models as \refig{fig:by_eta_eta}.}
    \label{fig:flux_rate_eta}
\end{figure}

%-------------------------------------------------------------------------------

\subsection{Comparison with the uniform resistivity case}
Our four models with constant and uniform resistivity profiles (\rmhd{10}{C2}{1} to \rmhd{10}{C5}{1}) show the profound impact that the adoption of effective resistivity has on the reconnection dynamics.
The magnetic dissipation $\eta_0$ ranges from  $2.5\times10^{-2}$ to $2.5\times10^{-5}$, with the corresponding macroscopic Lundquist number $S=Lc_A/\eta$ going from $\simeq37.8$ to $\simeq3.78\times10^4$.
All models were produced with the same resolution as the previous ones, i.e. $N_x=4096$.
% BY
For the highest values of resistivity we tested ($\eta=2.5\times10^{-2}$) the reconnected field $B_y$ experiences only a marginal initial growth, then decaying continuously after $t\sim5$ (top panel of \refig{fig:by_eta_eta}).
This is due to the fact that the dissipation of the initial magnetic field across the whole domain leads to a widening of the current sheet, therefore acting against its fragmentation into plasmoids.
For lower values of $\eta$ the reconnected field grows faster during the transient set by the initial perturbations, followed by a sudden rise once the fastest mode selected by the tearing instability becomes sufficiently strong.
This transition occurs earlier for lower resistivity, but it remains still very slow compared to the growth produced by models \rmhd{10}{E}{1} and \rmhd{10}{Ed0}{1 }.
Only the simulation with the lowest constant resistivity, i.e. \rmhd{10}{C5}{1}, shows a beginning of faster growth around $t\simeq4L_0/c$ on time scales compatible with the onset of the ideal tearing mode (i.e. $\simeq1.66\tau_A$), therefore still slower than the effective resistivity cases.
This simulation unfortunately crashed due to the low value of resistivity imposed, therefore it remains unclear to what extent such growth would continue over time. 

% ETA + FLUXES + RATE
Despite the average resistivity within the current sheet of models \rmhd{10}{E}{1} and  \rmhd{10}{Ed0}{1} falls within the range sample by our simulations with constant $\eta$ (bottom panel of \refig{fig:by_eta_eta}), none of latter models leads to a reconnection dynamics that is as fast as in the PIC case.
For all simulations with constant resistivity the reconnected flux $\Phi_{\rm rec}$ grows much more slowly than the effective resistivity case (top panel of \refig{fig:flux_rate_eta}) and produce reconnection rates that are at least one order of magnitude smaller than PIC or effective resistivity cases (bottom panel). 

The sole exception to this trend is model \rmhd{10}{C5}{1}, which starts producing a significant increase of $\Phi_{\rm rec}$ around $t\simeq4$.
This is consistent with the growth of $B_y$ for this simulation, which coincides also with a faster reconnection rate peaking at $\sim0.05$.
This is due to the fact that the resistivity is approaching the threshold for the onset of the ideal tearing instability, which in our case corresponds to $\eta_{\rm ideal}\simeq0.945\times10^{-6}$ for $S=10^6$, $\sigma_0=10$, $\beta_0=0.01$, and $a/L=100$.
However, the ideal tearing growth rate is still only $60\%$ of the inverse Alfv\'en crossing time $\tau_A^{-1}$, while for models with effective resistivity the instability develops on a timescale $\tau_A$. 
It also remains unclear (given the limited duration of the simulation) to what extent the reconnection rate might remain over time significantly higher than cases with higher resistivity.

Our simulations with constant resistivity allow us to estimate the numerical resistivity for a grid spacing $\Delta x\simeq0.002$, that is $\eta_{\rm num}\lesssim10^{-5}$.
This can be seen from the fact that model \texttt{s10eC5r1} shows a qualitatively different behavior than the cases with higher resistivity, i.e. the onset of the ideal tearing instability.
This estimate seems also to be consistent with Eq.~101 from \cite{komissarov2024} 
\begin{equation}
    \eta_{\rm num} = A_\eta\frac{\Delta x}{\Delta t}\mathcal{L}\left(\frac{\Delta x}{\mathcal{L}}\right)^r\simeq 10^{-5}\; ,
\end{equation}
where we used grid spacing $\Delta x=0.002$, characteristic length $\mathcal{L}=1$, time step $\Delta t=10^{-4}$, order $r=2$, and normalization factor $A_\eta=0.1$ \citep[which is a conservative estimate based on the normalization reported by][]{komissarov2024}.
%-------------------------------------------------------------------------------

%%%%%%%%%%%%%%%%%%%%%%%%%%%%%%%%%%%%%%%%%%%%%%%%%%%%%%%%%%%%%%%%%%%%%%%%%%%%%%%%
%%%%%%%%%%%%%%%%%%%%%%%%%%%%%%%%%%%%%%%%%%%%%%%%%%%%%%%%%%%%%%%%%%%%%%%%%%%%%%%%
%%%%%%%%%%%%%%%%%%%%%%%%%%%%%%%%%%%%%%%%%%%%%%%%%%%%%%%%%%%%%%%%%%%%%%%%%%%%%%%%
%%%%%%%%%%%%%%%%%%%%%%%%%%%%%%%%%%%%%%%%%%%%%%%%%%%%%%%%%%%%%%%%%%%%%%%%%%%%%%%%
%%%%%%%%%%%%%%%%%%%%%%%%%%%%%%%%%%%%%%%%%%%%%%%%%%%%%%%%%%%%%%%%%%%%%%%%%%%%%%%%

\section{Impact of numerical and physical parameters}\label{sec:parameters}
\subsection{Grid resolution}\label{subsec:res}
We explored the impact of grid resolution on our effective resistivity models by reproducing model \rmhd{10}{E}{1} using different numbers of grid points $N_x=\{256,512,1024,2048,4096\}$.
As we keep all other parameters fixed, these models resolve the upstream plasma skin depth respectively with 1.25, 2.5, 5, 10 and 20 grid points.

\begin{figure}
    \centering
    \includegraphics[width=.48\textwidth]{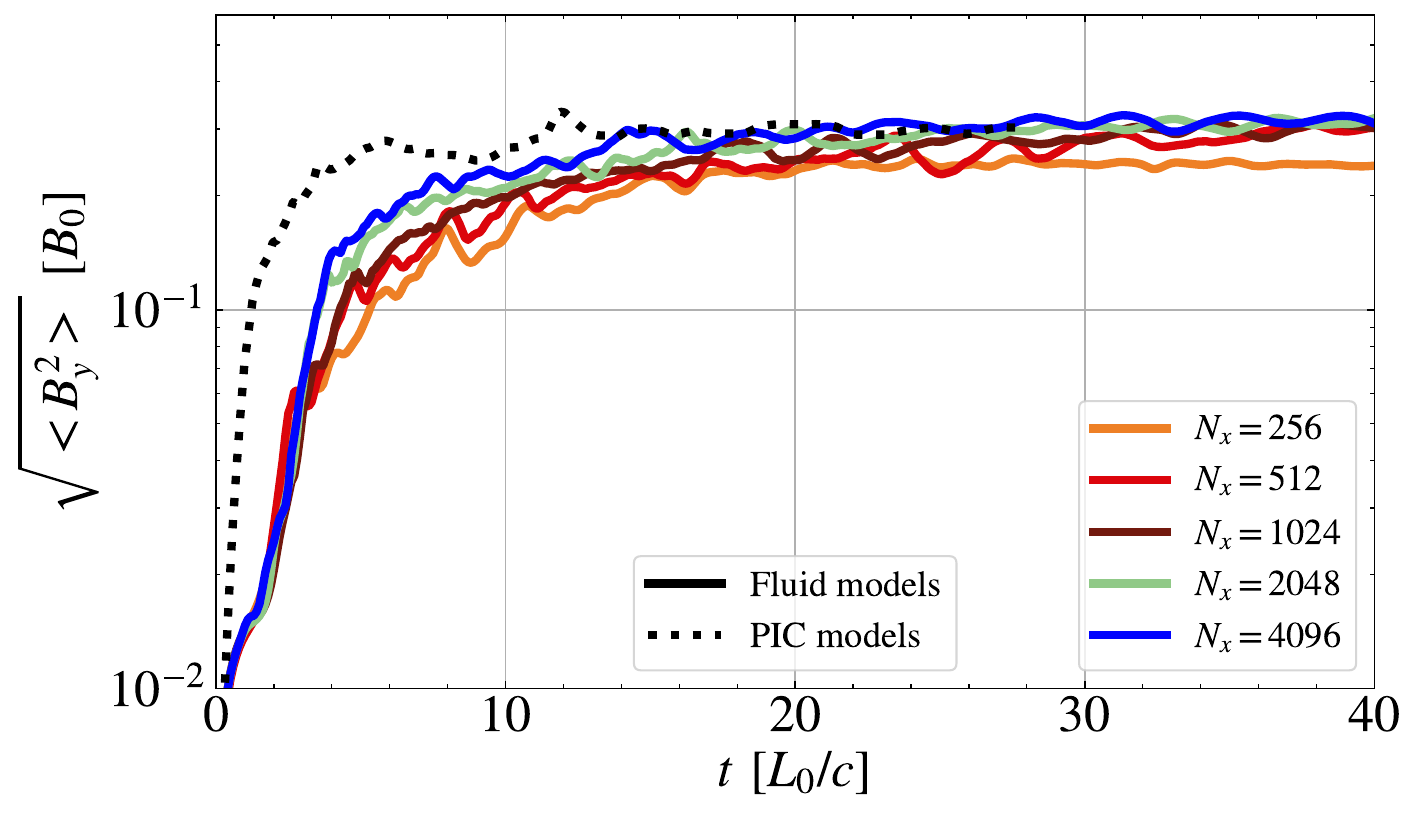}
    \includegraphics[width=.48\textwidth]{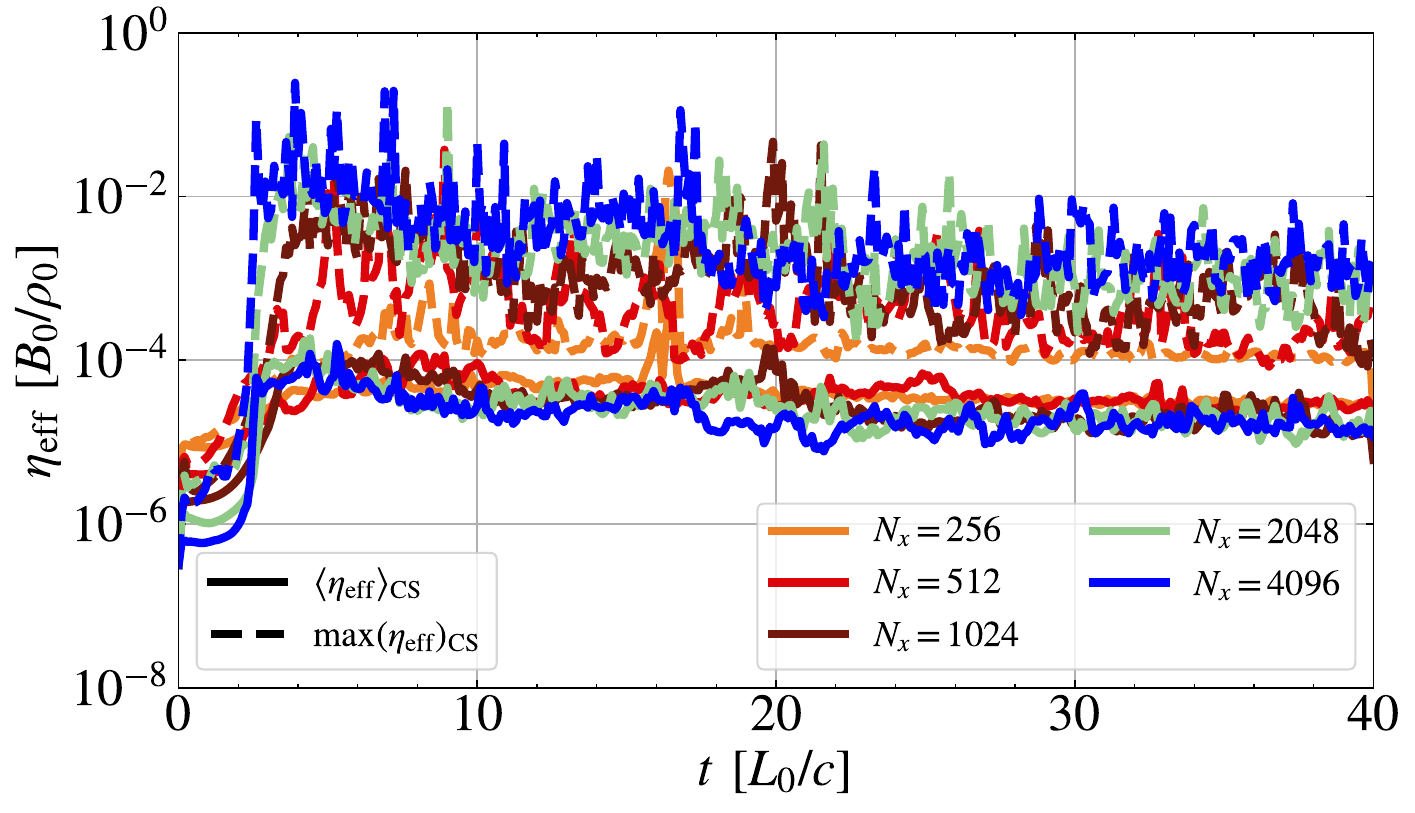}
    \caption{Top: RMS value of the transverse component of the magnetic field over time for model \rmhd{10}{E}{1} reproduced with different resolutions.
    Bottom: average and maximum value of the resistivity over time.
    The average is calculated over a rectangular box of sizes $L_x\times 2a$ centered in $y=0$.}
    \label{fig:by_eta_res}  
\end{figure}

From the top panel of \refig{fig:by_eta_res} we can see that already with $N_x=2048$ we obtain fully converged results for the growth and saturation of the RMS $y$-component of the magnetic field.
The development of the tearing instability before its saturation between $t\simeq5$ and $t\simeq10$ is quantitatively well captured already with $N_x=1024$ points.
Even at very modest resolutions ($N_x=256$) the effective resistivity prescription is capable to capture the fast onset of magnetic reconnection (although the field is never able to reach the same saturation level by the end of the simulation).
In contrast, models with constant resistivity approaching the fast reconnection regime require generally much higher resolutions to reach convergence \citep{puzzoni2021}.
The fact that a coarser resolution leads to lower values of the reconnected magnetic field at saturation stems mainly from two effects. 
The increased numerical dissipation leads, on the one hand, to a stronger diffusion of $B_x$ at the current sheet interface with the external plasma, reducing the magnetic field reservoir available for reconnection and opposing the thinning of the current sheet.
On the other hand, a coarser resolution is also associated to a lower effective resistivity within the current sheet during its fragmentation (see bottom panel of \refig{fig:by_eta_res} between $t\simeq2.5$ and $t\simeq5$) where, despite still being larger than the numerical one, still produces a weaker reconnection event. 

\begin{figure}
    \centering
    \includegraphics[width=.48\textwidth]{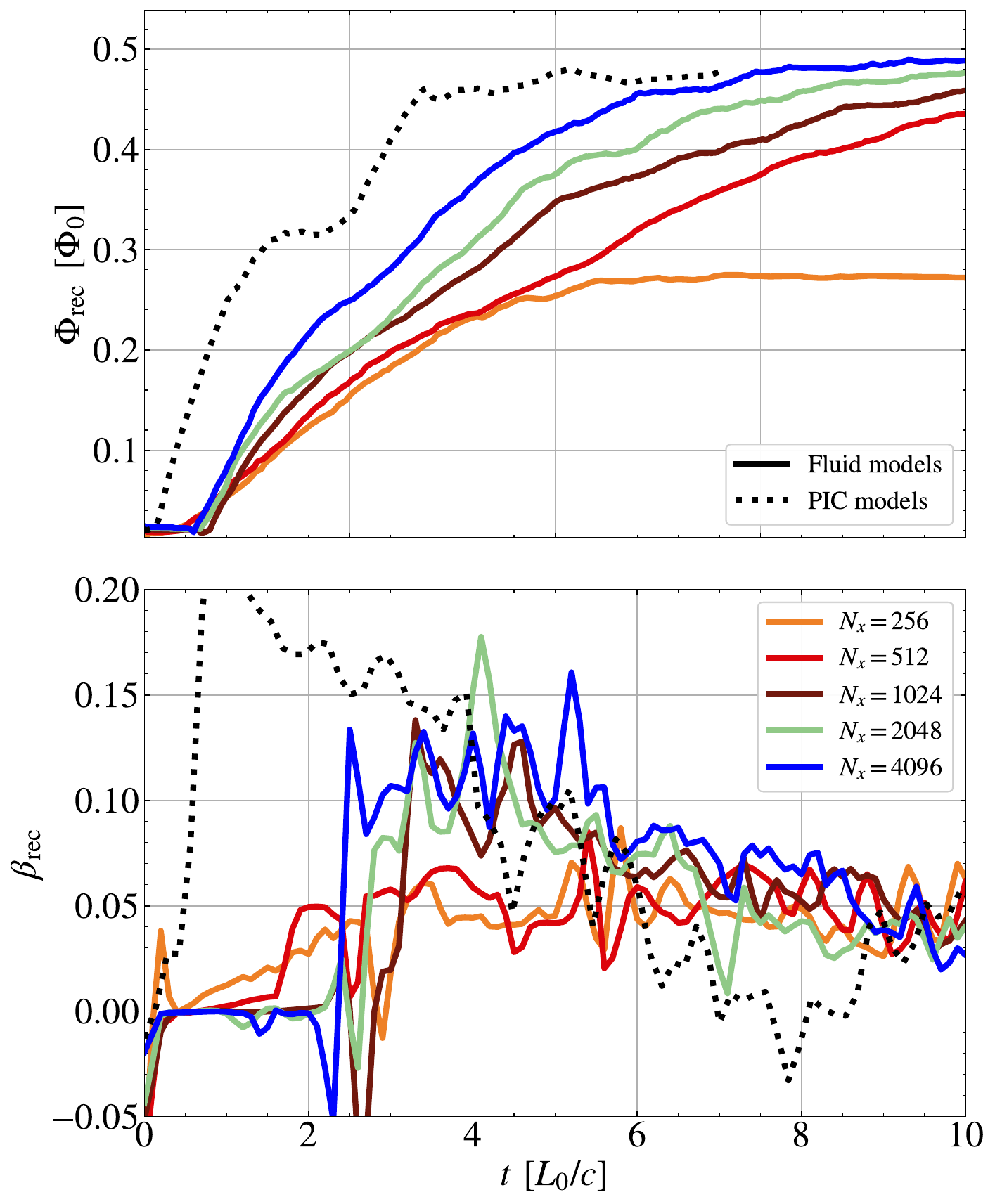}
    \caption{Reconnected magnetic flux (top) and reconnection rate (bottom) over time for the same models as \refig{fig:by_eta_res}.}
    \label{fig:flux_rate_res}
\end{figure}

The reconnecting magnetic flux and the associated reconnection rate shown in \refig{fig:flux_rate_res} provide a more detailed picture.
Results obtained with $N_x\gtrsim2000$ are quantitatively in agreement in terms of the amount of magnetic field reconnected by the $t=40 L_u/c$ mark and the corresponding reconnection rate.  
However, for $N_x=4096$ we obtain faster and more steady reconnection, reaching the PIC saturation level by $t\simeq30$.
At the lowest resolution ($N_x=256$) reconnection reaches a halt after the initial fragmentation and merging of the first plasmoids around $t\simeq20 L_u/c$, at which point the two last magnetic islands remain unmerged for the rest of the simulations and the reconnected magnetic flux stalls.
At twice the resolution ($N_x=512$) the increase in reconnected flux in the first half of the simulation is not significantly impacted, as shown also by the similar reconnection rate of $\simeq 0.05$.
However, with 512 points along the $x$-direction the model is able to continue the merging of the last large magnetic islands and approach the expected saturation level.
Overall, the resulting reconnection rates are qualitatively similar among models at different resolutions, with simulations running on the coarser grids showing a decrease by at most a factor $\sim2$ in the first stages of the reconnection event.

%-------------------------------------------------------------------------------

\subsection{Initial upstream density}\label{subsec:rho}
The effective resistivity expressed in \refeq{eq:eff_eta5} is completely defined by the dynamic state of the plasma and the characteristic length scale $\bar{\delta}_u$.
However, the initial choice of upstream density $\rho_0$ can in principle have a direct impact on the properties of magnetic dissipation, since the effective resistivity scales as $\eta_{\rm eff}\propto \rho^{-1}$.  
To quantify the impact of this dependence, we reproduced model \rmhd{10}{E}{1} with a value of $\rho_0$ decreased and increased by a factor 10 (models \rmhd{10}{E}{10} and \rmhd{10}{E}{01}, respectively).
Changing this parameter modifies not only the magnitude of the magnetic dissipation, but also the intensity of the upstream magnetic field $B_0=\sqrt{\rho_0\sigma_0}$ (for a fixed magnetization).

\begin{figure}
    \centering
    \includegraphics[width=.48\textwidth]{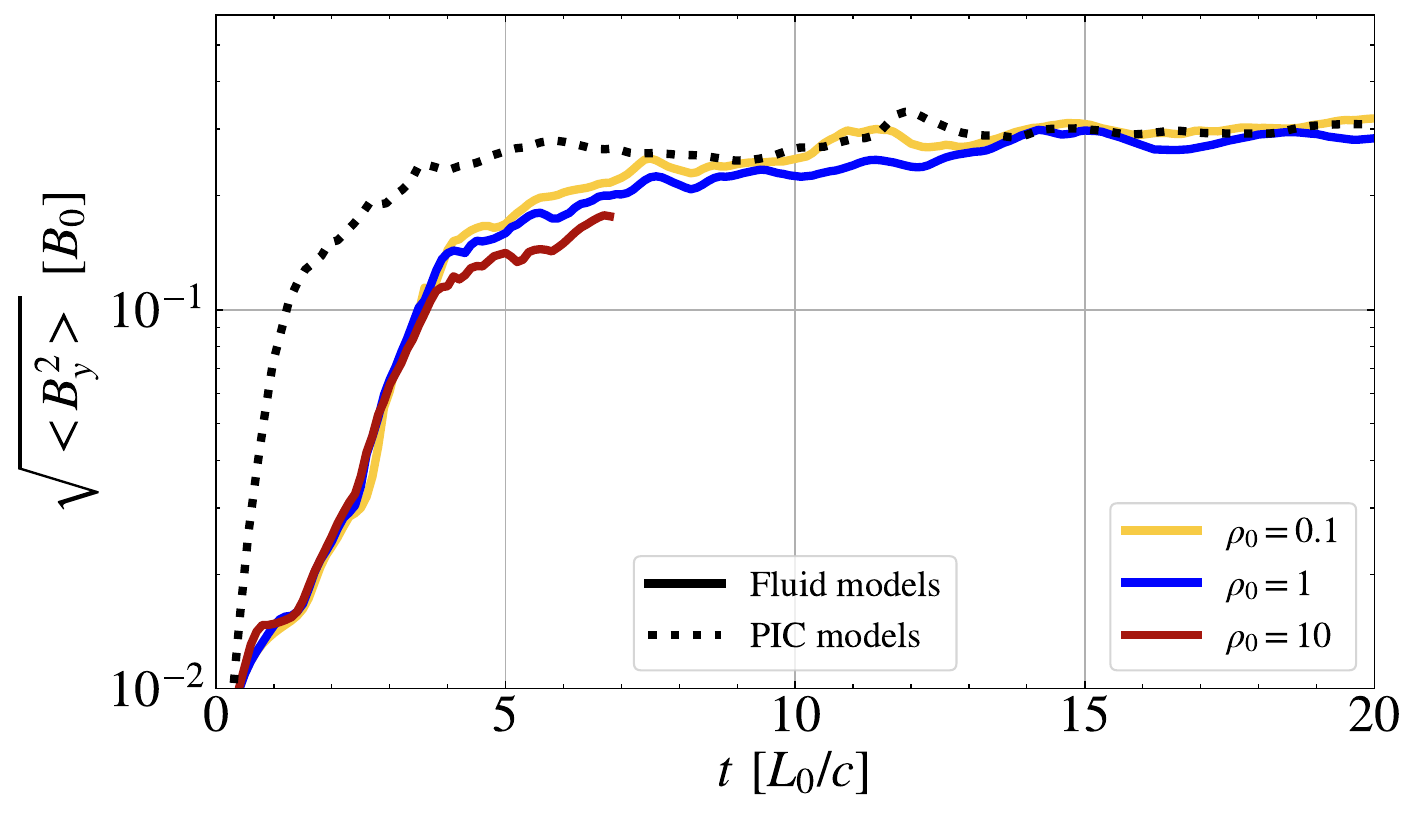}
    \includegraphics[width=.48\textwidth]{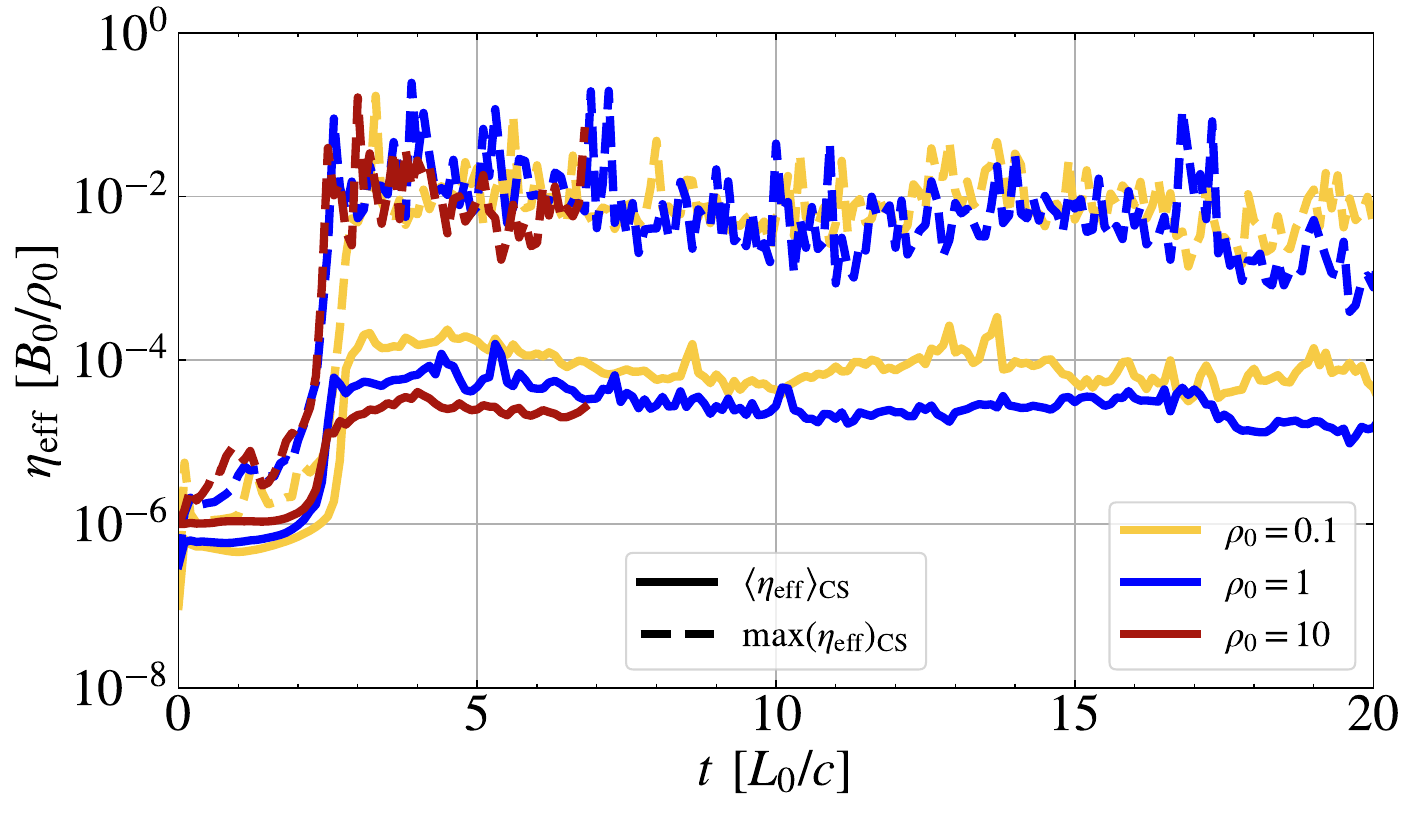}
    \caption{Top: RMS value of the transverse component of the magnetic field over time for models \rmhd{10}{E}{10}, \rmhd{10}{E}{1}, and \rmhd{10}{E}{01}.
    Bottom: average and maximum value of the resistivity over time.
    The average is calculated over a rectangular box of sizes $L_x\times 2a$ centered in $y=0$.}
    \label{fig:by_eta_rho}  
\end{figure}

\begin{figure}
    \centering
    \includegraphics[width=.48\textwidth]{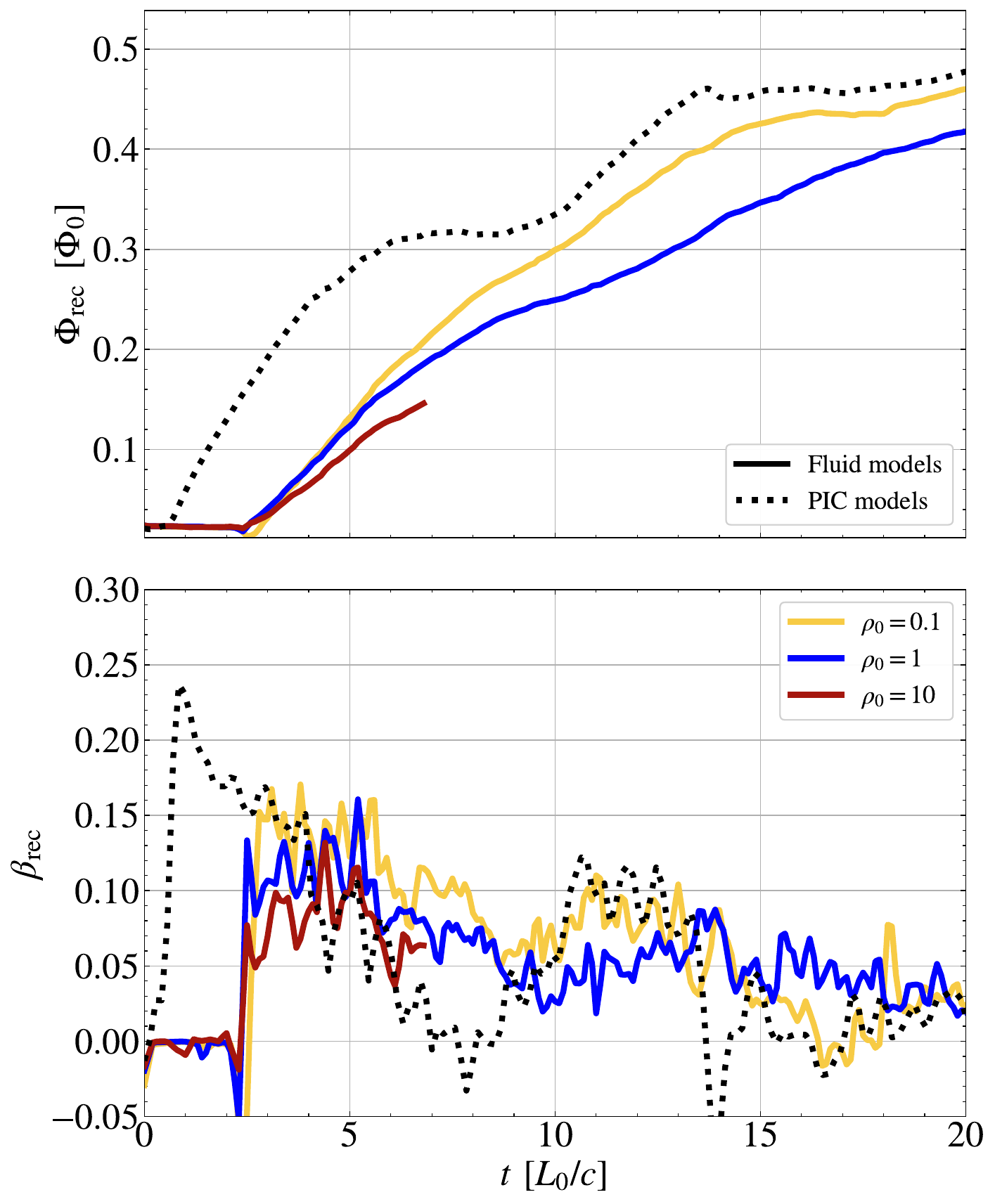}
    \caption{Reconnected magnetic flux (top) and reconnection rate (bottom) over time for the same models as \refig{fig:by_eta_rho}.}
    \label{fig:flux_rate_rho}
\end{figure}

The evolution of the RMS reconnected magnetic field $B_y$ (measured in units of $B_0$) appears to be rather insensitive to the variation in $\rho_0$ during the initial exponential growth phase (top panel of \refig{fig:by_eta_rho}).
This is consistent with the fact that the upstream Alfv\'en velocity in all these models is the same, since neither $\sigma_0$ nor $\beta_0$ have changed.
However, models with higher density tend to start the saturation phase earlier and lead to slightly less reconnected magnetic field.
Similarly to model \rmhd{10}{C5}{1}, our simulation with high background density did not reach its conclusion, as a generally much lower value of the resistivity decreases the numerical stability of the code. 
The temporal profiles of the maximum value assumed by the effective resistivity (bottom panel of \refig{fig:by_eta_rho}) show how this quantity scales indeed with $B_0/\rho_0$.
The average resistivity along the current sheet shows instead some deviations from such scaling behavior, with again higher densities being associated with even lower resistivities.
A similar dependence on $\rho_0$ can be found by looking at the reconnected magnetic flux (\refig{fig:flux_rate_rho}), with least and most dense cases delivering the fastest and slowest growth of $\Phi_{\rm rec}$, respectively.
While the differences between the two less dense models (\rmhd{10}{E}{01} and \rmhd{10}{Ed0}{1}) do not exceed the level of $10\%$ by the time the two last plasmoids are fully formed ($t\simeq7$), the highest density case presents a systematically slower increase of reconnected flux, which directly impacts the corresponding reconnection rates (bottom panel), with $\beta{\rm rec}$ ranging from about $\sim0.085$ (model \rmhd{10}{E}{10}) to $\simeq0.11$ (model \rmhd{10}{E}{01}).


%-------------------------------------------------------------------------------

\subsection{Initial magnetization}
\begin{figure}
    \centering
    \includegraphics[width=.48\textwidth]{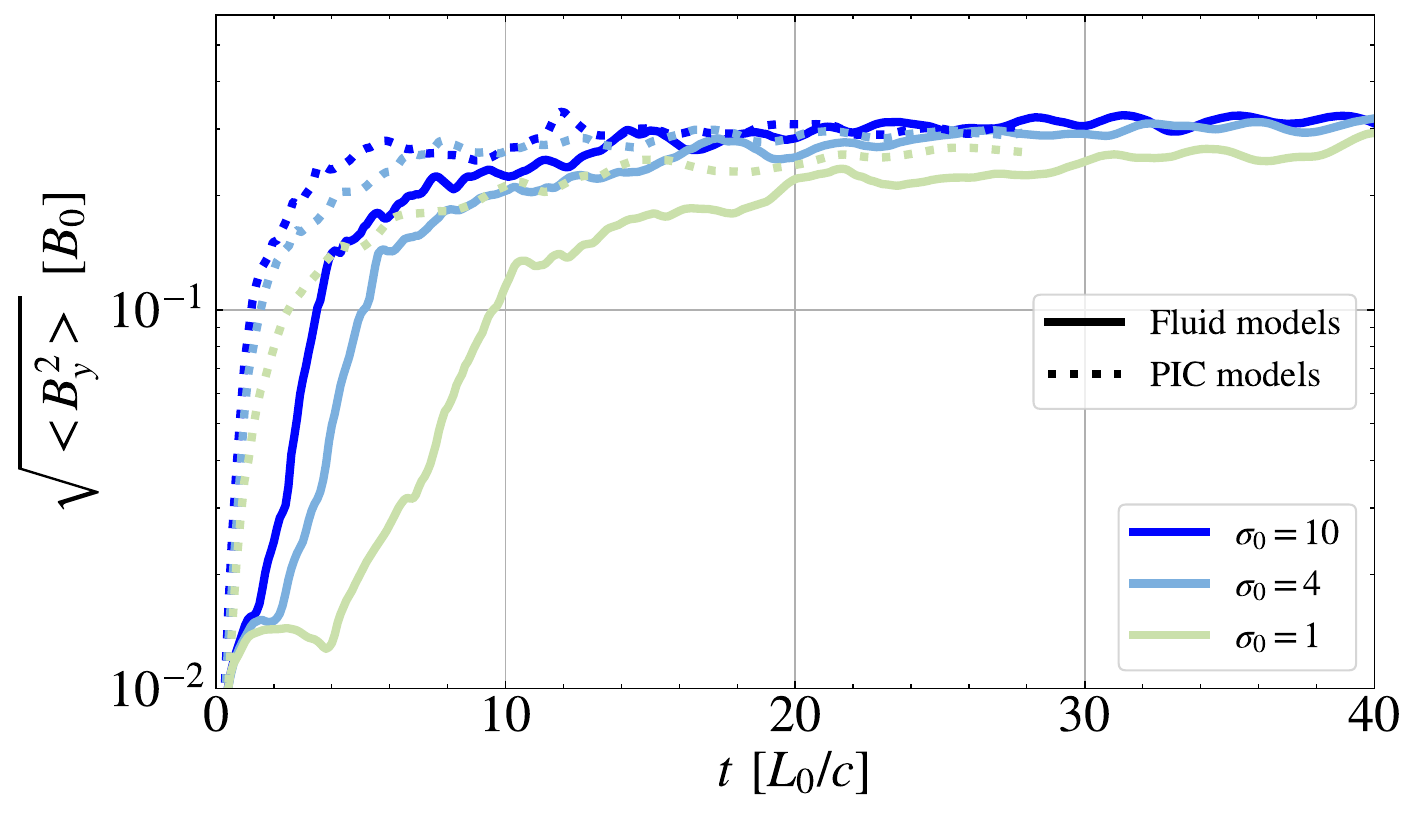}
    \includegraphics[width=.48\textwidth]{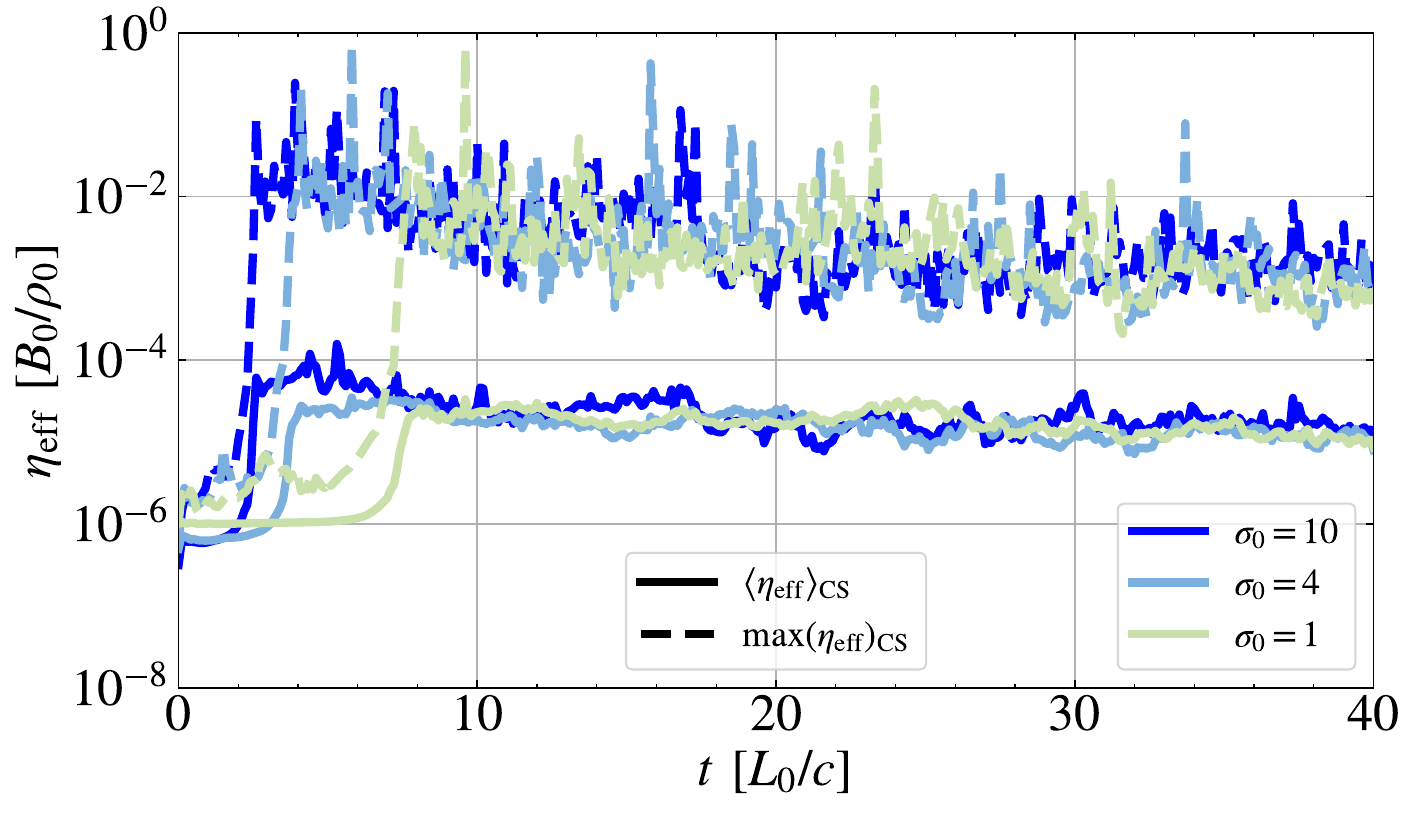}
    \caption{Top: RMS value of the transverse component of the magnetic field over time for models \rmhd{10}{E}{1}, \rmhd{4}{E}{1}, and \rmhd{1}{E}{1}.
    Bottom: average and maximum value of the resistivity over time.
    The average is calculated over a rectangular box of sizes $L_x\times 2a$ centered in $y=0$.}
    \label{fig:by_eta_sigma}  
\end{figure}

We now analyze the role played by the upstream magnetization $\sigma_0$ in determining the reconnection dynamics by considering two lower values of initial magnetization in addition to the previously discussed benchmark, i.e. $\sigma_0\in\{1,4,10\}$.
For these setups we produced ResRMHD models with effective resistivity as well as the corresponding PIC simulations.
Our chosen values for $\sigma_0$ imply an asymptotic magnetic field $B_0\in\{1,2,3.16\}$ and an upstream Alfv\'en velocity of $c_{A,0}\in\{0.887,0.704,0.945\}$, respectively.
Both kinetic and fluid models exhibit similar trends with varying magnetization, with a slower growth of the RMS $B_y$ field for lower values of $\sigma_0$ (\refig{fig:by_eta_sigma}).
This is consistent with the expectation that the dynamic time scale of reconnection in the linear phase of  the ideal tearing regime should scale with the Alfv\'en crossing time $\tau_A=L_u/c_{A,0}$ \citep{delzanna2016}.
A lower magnetization leads also to a lower saturation of the transverse magnetic field $B_y$, an effect that is more pronounced in model \rmhd{1}{E}{1}.
The evolution of the resistivity presents a similar dilation of the dynamic time scale of the reconnection event (bottom panel of \refig{fig:by_eta_sigma}).
Both the average and maximum values of $\eta_{\rm eff}$ are also consistent among different models once they are normalized by $B_0/\rho_0$, which is a further confirmation of the scaling behavior of $\eta_{\rm eff}$ with the upstream magnetic field and plasma density. 

\begin{figure}
    \centering
    \includegraphics[width=.48\textwidth]{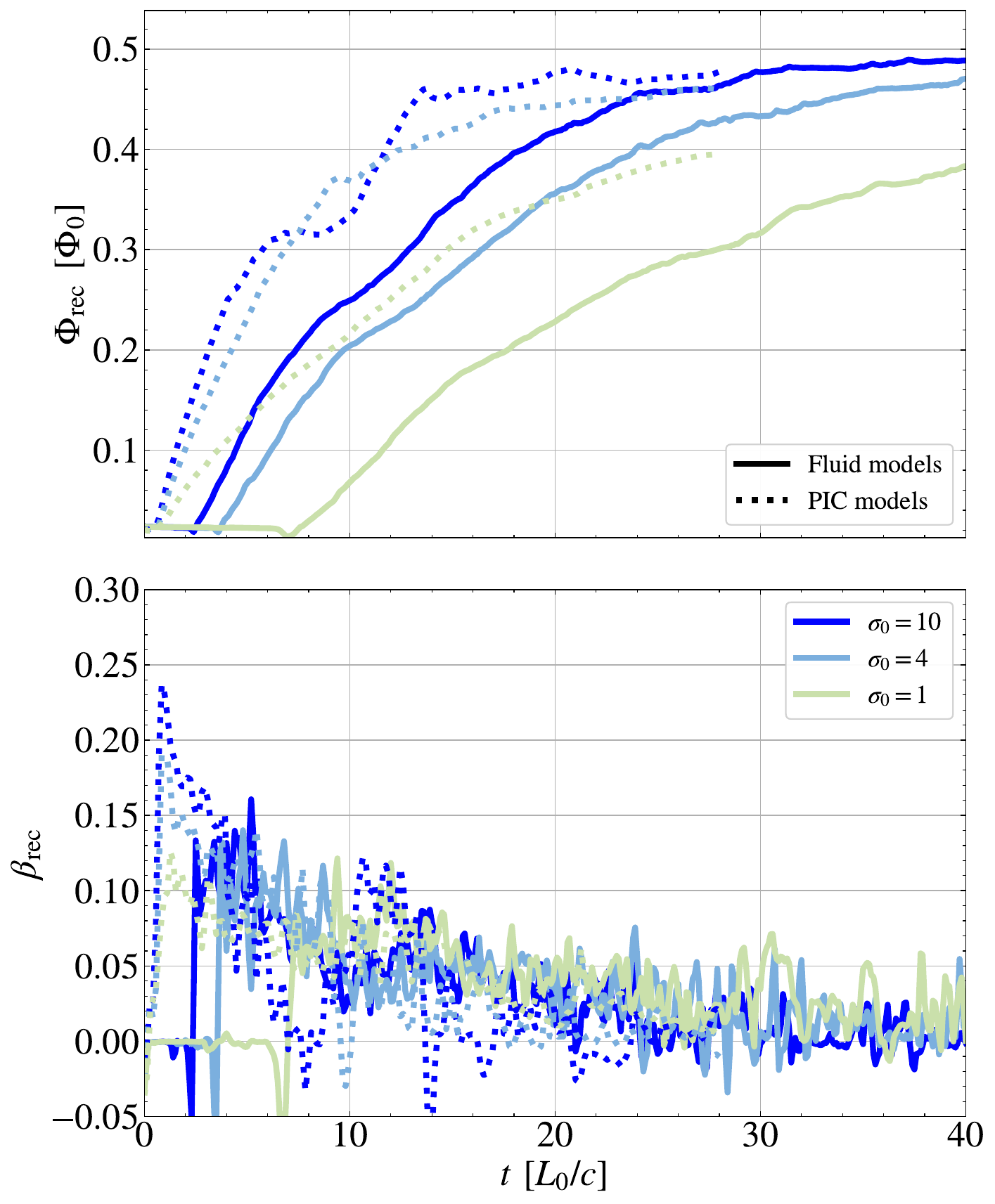}
    \caption{Reconnected magnetic flux (top) and reconnection rate (bottom) over time for the same models as \refig{fig:by_eta_sigma}.}
    \label{fig:flux_rate_sigma}
\end{figure}

The evolution over time of the reconnected magnetic flux in \refig{fig:flux_rate_sigma} clearly shows that our ResRMHD models have a systematic initial delay due to the initial linear phase of the tearing instability, whose duration increases for lower magnetizations. 
This is reflected in the current sheet fragmentation into smaller plasmoids occurring at later times for smaller values of $\sigma_0$.
PIC models do not experience any such delay for decreasing magnetizations, but have a similar decrease in the growth of reconnected flux.
The corresponding reconnection rates confirm that our ResRMHD models with effective resistivity lead to reconnection events with efficiency in quantitative agreement with PIC models, corroborating the results presented in Section~\ref{sec:general_analysis}.
Moreover, the reconnection rate after the initial transient does not appear to dependent too much on the magnetization (bottom panel of \refig{fig:flux_rate_sigma}), which is expected when one considers that lower values of sigma lead not only to a slower increase of the reconnected flux, but also lower Alfv\'en speeds.

%-------------------------------------------------------------------------------

\subsection{Reconnection rate trends}
\begin{figure}
    \centering
    \includegraphics[width=.48\textwidth]{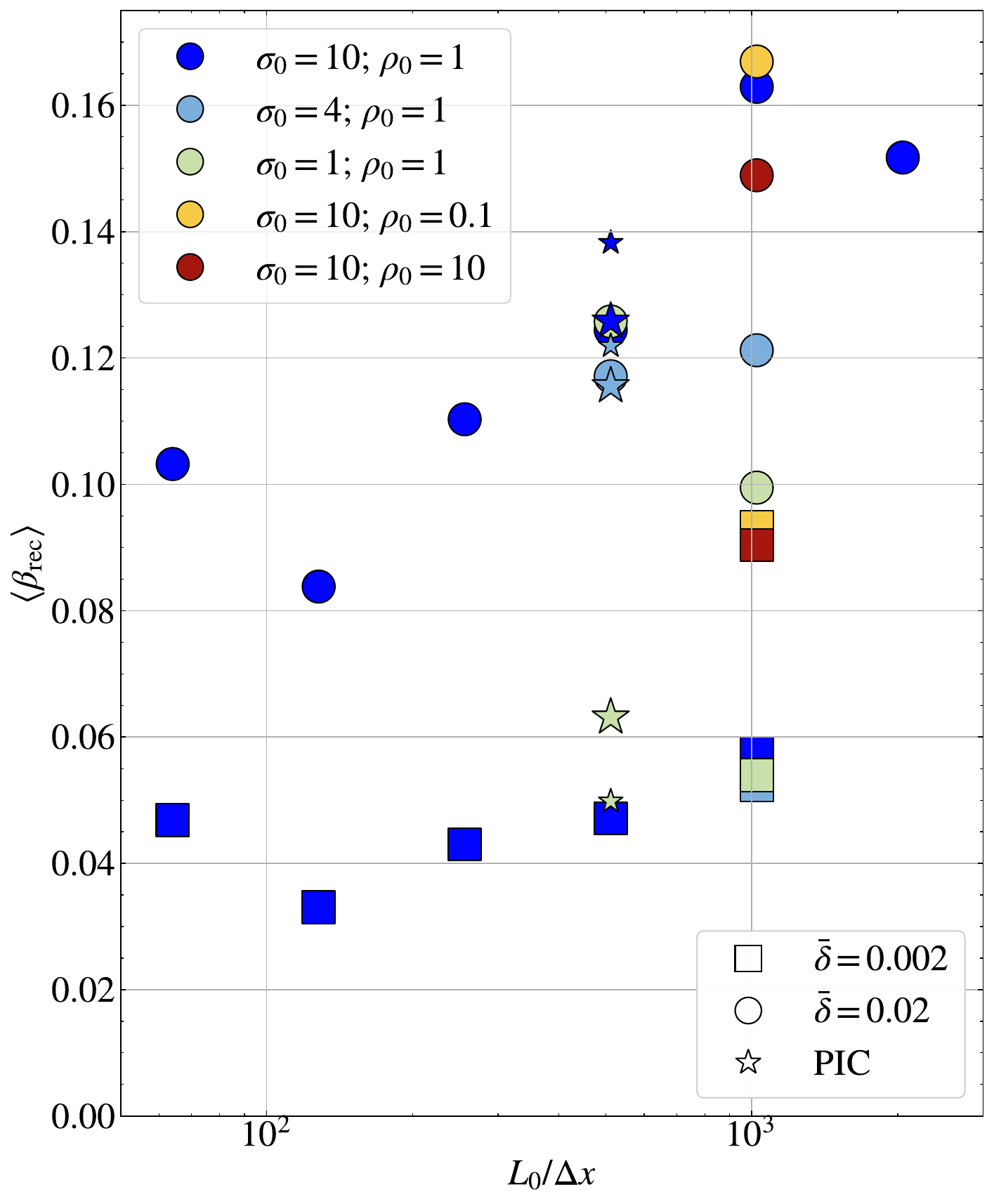}
    \caption{Average reconnection rate against resolution for the ResRMHD models with effective resistivity (circles and squares) and the PIC models (stars).
    Small and large stars refer to PIC models having initial perturbations with amplitude $0.001$ and $0.05$, respectively. 
    }
    \label{fig:rec_rate_diagram}
\end{figure}

We now compare more quantitatively the reconnection rates obtained in this work by computing a representative time average for each one of simulations we performed.
To exclude the initial transient and the later stage close to the saturation, we calculate the average in the time interval where the reconnected magnetic flux is between 30\% and 60\% of its maximum value.
We apply this procedure to a larger set of simulation with respect to the ResRMHD and PIC models presented in the previous sections, which leads to the scatter plot of $\langle\beta_{\rm rec}\rangle$ against spatial resolution in \refig{fig:rec_rate_diagram}.


The dependence on grid resolution is evident for both models \rmhd{10}{E}{1} and \rmhd{10}{Ed0}{1}, as the reconnection rate increases by up to a factor $\sim1.5$ between the most coarse grid and the reference cases discussed in Section \ref{sec:general_analysis}.
A run with $N_x=8192$ for model \rmhd{10}{Ed0}{1} suggests, however, that there should be no further quantitative deviations for higher resolutions and that indeed our measured rates are numerically converged.
All fluid models produce reconnection rates within the same order of magnitude of the kinetic simulations, with the limit case of $\bar{\delta}=0.02$ being in quantitative agreement with the corresponding PIC realizations.
We assessed the variance of the rates produced by the kinetic models by performing an additional set of simulations with the same range of values for $\sigma_0$ but weaker initial perturbations ($\epsilon=0.001$ rather than $0.05$), which are represented by the smaller stars in \refig{fig:rec_rate_diagram} and deviate from the runs with stronger perturbation by 10-20\% at most.
The same trend with magnetization is reproduced by all PIC and ResRMHD models, i.e. there is little variance for different values of $\sigma_0$.
Simulations with $\bar{\delta}=0.02$ and $N_x=4096$ appear to be an exception, with the reconnection rate increasing by up to 60\% with higher magnetizations.
The PIC models with $\sigma_0=1$ also deviate from the other kinetic simulations, this time by a factor $\sim2$. 
Changing the upstream plasma density impacts the corresponding reconnection rate for both values of the plasma skin depth parameter explored.
A variation over two orders of magnitude of $\rho_0$ introduces a deviation in the average reconnection rate of $\Delta\langle\beta_{\rm rec}\rangle$ between $\sim0.02$ and $\sim0.04$.
Model \rmhd{10}{E}{10} (red square) appears to have larger $\langle\beta_{\rm rec}\rangle$ than expected from the discussion in Section \ref{subsec:rho}. 
This is likely an artifact due to the short duration of the simulation, which prevents a more significant decrease in the reconnection rate over time.

%%%%%%%%%%%%%%%%%%%%%%%%%%%%%%%%%%%%%%%%%%%%%%%%%%%%%%%%%%%%%%%%%%%%%%%%%%%%%%%%
%%%%%%%%%%%%%%%%%%%%%%%%%%%%%%%%%%%%%%%%%%%%%%%%%%%%%%%%%%%%%%%%%%%%%%%%%%%%%%%%
%%%%%%%%%%%%%%%%%%%%%%%%%%%%%%%%%%%%%%%%%%%%%%%%%%%%%%%%%%%%%%%%%%%%%%%%%%%%%%%%
%%%%%%%%%%%%%%%%%%%%%%%%%%%%%%%%%%%%%%%%%%%%%%%%%%%%%%%%%%%%%%%%%%%%%%%%%%%%%%%%
%%%%%%%%%%%%%%%%%%%%%%%%%%%%%%%%%%%%%%%%%%%%%%%%%%%%%%%%%%%%%%%%%%%%%%%%%%%%%%%%

\section{Conclusions}\label{sec:conclusions}
We presented a series of ResRMHD numerical simulations of magnetic reconnection that make use, for the first time, of an effective resistivity formulation derived from recent first-principle PIC models, thus bridging the gap between fluid and kinetic models of astrophysical plasmas.
We adapted the closure obtained by \cite{selvi2023} to express the resistivity in terms of fluid quantities and known constants, which allows one to constrain the dissipation properties of the plasma by its local state and the characteristic length scales of the reconnecting current sheet.

% General results
All our ResRMHD models with effective resistivity lead to a fast reconnection of the initial magnetic field, with a quick fragmentation of the current sheet into plasmoids and the saturation into a large magnetic island within $\simeq$ 20 dynamical timescales.
The evolution of the current sheet is in good agreement between fluid and kinetic models, which confirms the capability of the effective resistivity to capture the intrinsically collisionless nature of magnetic reconnection.
To capture the onset of the ideal tearing instability on dynamical timescales, constant resistivity models require very low values of $\eta$, and therefore sufficiently high grid resolutions to describe the smaller dissipation length scale of the simulation.
On the other hand, the effective resistivity formulation allows one to retrieve fast reconnection even at very modest resolutions.
This is due to the fact that the enhanced magnetic dissipation within the current sheet induces strong reconnection without requiring the tearing instability to thin the current sheet's width.
Future works will compare more thoroughly the effective resistivity models with the dynamics of relativistic reconnection at high Lundquist numbers, which will require higher resolutions and will benefit from recently developed higher order schemes for RMHD and ResRMHD models \cite{berta2024,mignone2024}.  

% Main contribution from the electric field 
While both strong non-ideal electric fields and velocity gradients a priori can contribute to enhance the resistivity (see \refeq{eq:eff_eta5}), the former contribution dominates in the regions with the highest dissipation, as it is first-order in the dimensionless plasma skin-depth parameter $\bar{\delta}_u$ rather than second.
This suggests that our effective resistivity formulation might be simplified even more to include exclusively the non-ideal electric field, especially when applied to astrophysical systems with larger size than an isolated current sheet (e.g. a relativistic jet or a black hole magnetosphere).
Such formulation would be similar to the one presented in \cite{ripperda2019a}, where the resistivity was taken to be proportional to the local current density, but with the coupling constant ($\bar{\delta}$) representing the ``effective'' plasma skin depth that the fluid system is allowed to see.

The value of this coupling constant can have a significant impact on the resulting reconnection dynamics.
Taking $\bar{\delta}$ equal to the plasma skin depth in the current sheet leads to a fragmentation into plasmoids and merging phase that qualitatively resembles both PIC models and constant resistivity ResRMHD simulations.
The corresponding reconnection rates are much higher than the case with constant magnetic dissipation, but slightly lower than the kinetic counterparts.
On the other hand, a ten times larger plasma skin depth matching the upstream conditions of the problem reproduces more quantitatively the diagnostics obtained by first-principles simulations, but the underlying plasmoid dynamics deviates qualitatively from the standard scenario, as the increased dissipation leads to the formation of hot, rarefied cavities between large magnetic islands that prevent the production of smaller plasmoids connected by thinning reconnecting regions.
These differences are likely to have a significant impact on how particles are accelerated during the reconnection event, which we will assess in an upcoming work.
The choice of the numerical value for the $\bar{\delta}_u$ parameter becomes less evident for problems where the current sheet is not already present from the beginning. 
In our reconnection models we are in practice imposing a characteristic mass density unit that makes this parameter equal to the actual current sheet skin depth set by the corresponding PIC model.
Since a different value of $\bar{\delta}_u$ is likely to affect the formation of current sheets themselves (e.g. during the merge of two flux tubes), setting its value a priori requires more dedicated comparative studies between PIC and ResRMHD models. 
Future work will have to explore more thoroughly the dependence of our effective resistivity model on $\bar{\delta}_u$ for different initial pressure-balanced equilibria (e.g. with constant density or different temperatures), as well as for merging flux tubes \citep{ripperda2019b}.

% Impact of density and magnetization
Our simulations show that a higher background density tends to weaken the reconnection and decrease the reconnection rate, as the mean resistivity within the current sheet lowers by a factor of a few.
This is a direct consequence of the prescription we used for $\eta_{\rm eff}$ being inversely proportional to the plasma density.
We also verified that lower magnetizations produce slower reconnection events and lower reconnection rates (which is also expected in the constant resistivity case), with the mean resistivity scaling with the quantity $B_0/\rho_0$.

% Reconnection rate
The fast reconnection of magnetic flux translates occurring in our ResRMHD simulations is reflected by high reconnection rates (between $\sim0.05$ and $\sim0.2$).
These measurements are not only an order of magnitude larger than those produced in the constant resistivity case, but also in quantitative agreement with the corresponding PIC models.
This result showcases the potential of the effective resistivity formalism to reproduce within the standard ResRMHD framework the results of first-principles kinetic models, thus enabling a more accurate modeling of large-scale astrophysical systems where reconnection is expected to occur. 
Accretion flows on black holes would be a prime example of such an application, in particular when in the so-called MAD state \cite{tchekhovskoy2011}.
Since the amount of dissipation regulates the magnetic flux available at the event horizon, the inclusion of an effective resistivity might change the properties of the saturated state.
Moreover, a faster reconnection dynamics could profoundly impact the formation of possible particle acceleration sites in relativistic jets launched by newly formed compact objects \citep{mattia2023,mattia2024}. 
However, it should be noted that the prospects of resolving the actual skin depth of current sheets in large-scale astrophysical models are quite poor, given the large scale separation involved.
The application of our formulation to systems such as accreting compact objects or relativistic jets will require, therefore, to compromise for a much larger "effective" skin depth, since otherwise the resulting small resistivity would simply have no dynamical effects against the grid numerical dissipation (on top of leading to numerical instabilities).
This approach would be keen to the one adopted successfully by large scale GR-PIC models of relativistic magnetospheres \citep{crinquand2021,elmellah2022,crinquand2022}, which settle for a similar trade-off for the plasma skin depth.
Given this similarity, a closer comparison between such models and GRMHD simulations with effective resistivity might provide important insights on the dynamics of magnetic dissipation in a regime of strong gravity. 

% Strong approximations in deriving the expression for eta
Although our results show a remarkable agreement between fluid and kinetic models, it should be noted that the derivation of \refeq{eq:eff_eta5} relies on some strong assumptions.
In particular, we introduced the current density (and, therefore, the electric field) in the expression for $\eta_{\rm eff}$ to more directly tight the enhanced dissipation to the reconnection layer, but our derivation should be more rigorously verified to fully characterize the impact of non-collisional effects on magnetic dissipation within the ResRMHD framework.
Moreover, the ``kinetic'' formulation of the effective resistivity presented by \cite{selvi2023} has been obtained in the absence of a guide field and for a electron-positron plasma, while it remains unclear how the scenario would change with different magnetic field configurations or in the presence of ions.
Another limitation comes from the fact that the effective resistivity measurements were performed specifically in the X-points of the reconnecting current sheet.
Although they are undoubtedly the most relevant locations to study, further investigations with fully-kinetic models of relativistic reconnection are required to assess the general character of the proposed formulation.

Furthermore, our 2D results capture the main properties of how magnetic field is dissipated and particles are accelerated during a reconnection event, but they should be further tested in 3D to verify the extent of the validity of the prescription we used.
Finally, it should be noted that \refeq{eq:eff_eta1} and \refeq{eq:eff_eta5} assume the current sheet to be locally aligned with the $x$-axis, as the only velocity gradients appearing are those along the transverse direction $y$. 
This means that our prescription is likely to underestimate the contribution to the effective resistivity of the fluid velocity gradients, since the non-linear stage of the tearing instability leads to the formation of secondary current sheets perpendicular to the $x$-axis between merging plasmoids.
This aspect might be relevant when quantifying the efficiency of particle acceleration when using the effective resistivity formulation. 
A more consistent approach would require the identification of the local orientation of the current sheet to correctly include velocity gradients perpendicular to that direction.
We leave such improvement to our follow-up work.

%%%%%%%%%%%%%%%%%%%%%%%%%%%%%%%%%%%%%%%%%%%%%%%%%%%%%%%%%%%%%%%%%%%%%%%%%%%%%%%%
%%%%%%%%%%%%%%%%%%%%%%%%%%%%%%%%%%%%%%%%%%%%%%%%%%%%%%%%%%%%%%%%%%%%%%%%%%%%%%%%
%%%%%%%%%%%%%%%%%%%%%%%%%%%%%%%%%%%%%%%%%%%%%%%%%%%%%%%%%%%%%%%%%%%%%%%%%%%%%%%%
%%%%%%%%%%%%%%%%%%%%%%%%%%%%%%%%%%%%%%%%%%%%%%%%%%%%%%%%%%%%%%%%%%%%%%%%%%%%%%%%
%%%%%%%%%%%%%%%%%%%%%%%%%%%%%%%%%%%%%%%%%%%%%%%%%%%%%%%%%%%%%%%%%%%%%%%%%%%%%%%%

    \begin{acknowledgements}
        We thank the anonymous referee for improving this work with useful comments and suggestions.
        M.B. would like to thank Bart Ripperda, Sasha Philippov, Niccolò Bucciantini, and Emanuele Papini for useful discussions.
        This project has received funding from the European Union's Horizon Europe research and innovation programme under the Marie Sk\l{}odowska-Curie grant agreement No 101064953 (GR-PLUTO). 
        B.C. and E.F. are supported by the European Research Council (ERC) under the European Union’s Horizon 2020 research and innovation program (Grant Agreement No. 863412).
        L.D.Z. acknowledges support from the ICSC—Centro Nazionale di Ricerca in High-Performance Computing; Big Data and Quantum Computing, funded by European Union - NextGenerationEU.
        This work has received funding from the European High Performance Computing Joint Undertaking (JU) and Belgium, Czech Republic, France, Germany, Greece, Italy, Norway, and Spain under grant agreement No 101093441.
        This paper is also supported by the Fondazione ICSC, Spoke 3 Astrophysics and Cosmos Observations, National Recovery and Resilience Plan (Piano Nazionale di Ripresa e Resilienza, PNRR) Project ID CN\_00000013 ``Italian Research Center on High-Performance Computing, Big Data and Quantum Computing'' funded by MUR Missione 4 Componente 2 Investimento 1.4: Potenziamento strutture di ricerca e creazione di ``campioni nazionali di R\&S (M4C2-19)'' - Next Generation EU (NGEU).

        Computational resources on the OCCAM cluster (\url{https://c3s.unito.it/index.php/super-computer}) were provided by the Centro di Competenza sul Calcolo Scientifico (C3S) of the University of Torino.

        We also acknowledge CINECA for the availability of high performance computing resources and support on the LEONARDO supercomputer (\url{https://leonardo-supercomputer.cineca.eu/}) through a CINECA-INFN agreement, providing the allocations INF23\_teongrav and INF24\_teongrav. 
        Computing resources for performing the PIC simulations were provided by TGCC under the allocation A0150407669 made by GENCI.

        The figures presented in this paper were generated using the PyPLUTO (\url{https://github.com/GiMattia/PyPLUTO}) package (Mattia et al. in prep.) while all the line colors have been checked through a color blindness simulator (\url{https://www.color-blindness.com/coblis-color-blindness-simulator/}).
    \end{acknowledgements}

%%%%%%%%%%%%%%%%%%%%%%%%%%%%%%%%%%%%%%%%%%%%%%%%%%%%%%%%%%%%%%%%%%%%%%%%%%%%%%%%
%%%%%%%%%%%%%%%%%%%%%%%%%%%%%%%%%%%%%%%%%%%%%%%%%%%%%%%%%%%%%%%%%%%%%%%%%%%%%%%%
%%%%%%%%%%%%%%%%%%%%%%%%%%%%%%%%%%%%%%%%%%%%%%%%%%%%%%%%%%%%%%%%%%%%%%%%%%%%%%%%
%%%%%%%%%%%%%%%%%%%%%%%%%%%%%%%%%%%%%%%%%%%%%%%%%%%%%%%%%%%%%%%%%%%%%%%%%%%%%%%%
%%%%%%%%%%%%%%%%%%%%%%%%%%%%%%%%%%%%%%%%%%%%%%%%%%%%%%%%%%%%%%%%%%%%%%%%%%%%%%%%

    \bibliographystyle{aa} % style aa.bst
    \bibliography{main.bib}

    \begin{appendix}
    \section{Numerical benchmarks}
    We present here a series of tests designed to validate our implementation of the effective resistivity within the ResRMHD module of the \pluto{} code.
    \subsection{Uniform stationary resistive plasma}
    Using our prescription for the effective resistivity in \refeq{eq:eff_eta4} we can express Amp\`ere's law as
    \begin{equation} \label{eq:ampere_test1}
    	\partial_t \vec{E} - \vec{\nabla} \times \vec{B} = -\Gamma \frac{|J_0|}{\sqrt{E_0^2 + e_z^2}} [\vec{E} + \vec{v} \times \vec{B} - (\vec{E} \cdot \vec{v})\vec{v}] - q\vec{v}\; .
    \end{equation}
    If we now assume to have a stationary, unmagnetized, and uniform plasma with
    \begin{align*}
	\vec{v} = \vec{0};\ \ \ 
	\vec{B} = \vec{0};\ \ \ 
	P = 1;\ \ \ 
	\rho = 1\; ,
    \end{align*}
    we can reduce \refeq{eq:ampere_test1} to
    \begin{equation} \label{test1 eq.1}
	\frac {\partial \vec{E}} {\partial t} = -|J_0| \frac{\vec{E}}{|E_z|}\; .
    \end{equation}
    This equation requires only the integration of the $z$ component of the electric field, since we can take the other two as proportional to $E_z$ (i.e. $E_{x,y}=\alpha_{x,y}E_z$) and obtain the exact same differential equation for the electric field, namely
    \begin{equation}\label{eq:test1_diffeq}
	   \frac{dE_z}{dt} = -|J_0|\; \text{sign}(E_z)\; .
    \end{equation}
    The proportionality factors for $E_{x,y}$ are simply retrieved from the initial condition by requiring $E_{x,y}(t)/E_{x,y}(0)=E_z(t)/E_x(0)$.
    By integrating \refeq{eq:test1_diffeq} in time we finally obtain the analytical solution
    \begin{align*}
    	E_z(t) = 
        \begin{cases}
            E_z(0) - |J_0|\; \text{sign}(E_z) t &\text{for}\; t < \tau\\%[5pt]
            0 &\text{for}\ t \geq \tau\; ,
        \end{cases}
    \end{align*}
    where $\tau=E_z(0)/|J_0|$ is the timescale required for the electric field to be completely dissipated.
    Note that such solution differs substantially from the case with constant resistivity $\eta_0$, which would lead instead to an exponential decay of the electric field according to $\partial_t \vec{E} = -\eta_0^{-1} \vec{E}$.
    
\begin{figure}
        \centering
        \includegraphics[width=0.24\textwidth]{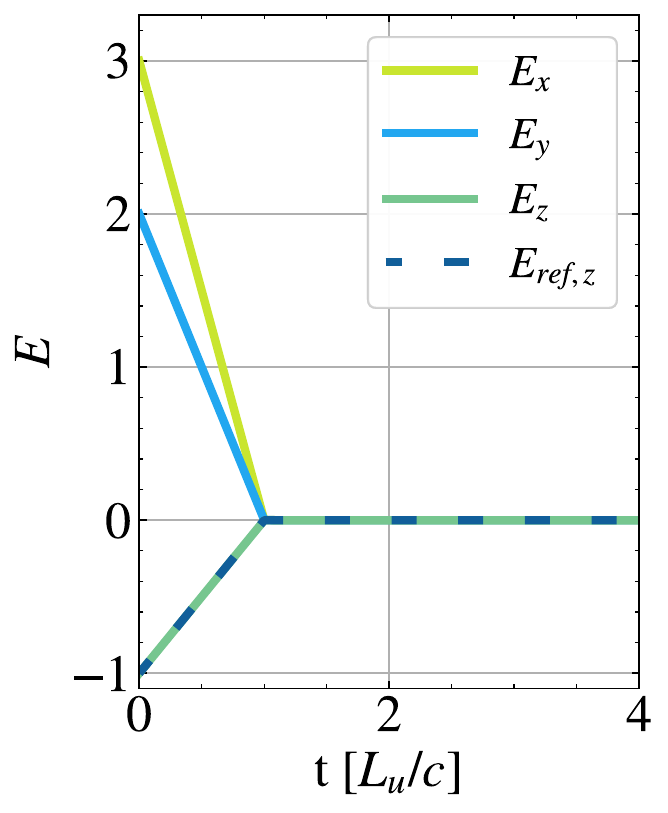}
        \includegraphics[width=0.24\textwidth]{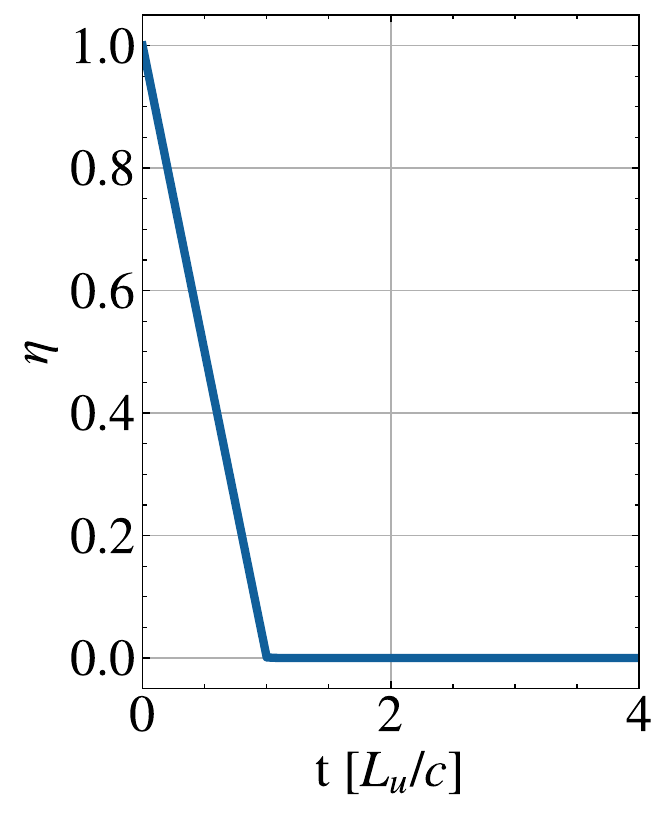}
        \caption{Electric field decay (left panel) and effective resistivity (right) over time for the uniform stationary resistive test with vanishing velocity gradients.
        The analytical solution is represented with the blue dashed line.}
        \label{fig:uniform_test1}
    \end{figure}

    The left panel of \refig{fig:uniform_test1} shows the results of a 1d simulation of such configuration using our effective resistivity prescription and an initial uniform electric field given by $\vec{E}=(3,2,-1)$ and setting $|J_0|=1$.
    The field correctly decays linearly with time, completely vanishing at $t=\tau=1$. 
    The effective resistivity starts with a value equal to $E_z/|J_0|=1$ and decays proportionally with the electric field, reaching the minimun floor value $\eta_{\rm floor}=10^{-6}$ at $t=1$ (right panel of \refig{fig:uniform_test1}).

    \begin{figure}
        \centering
        \includegraphics[width=0.24\textwidth]{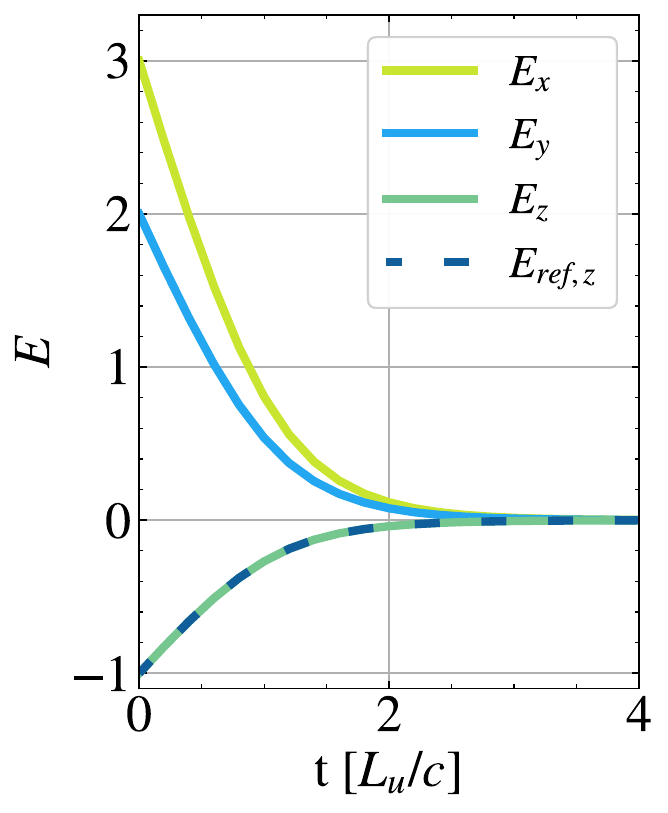}
        \includegraphics[width=0.24\textwidth]{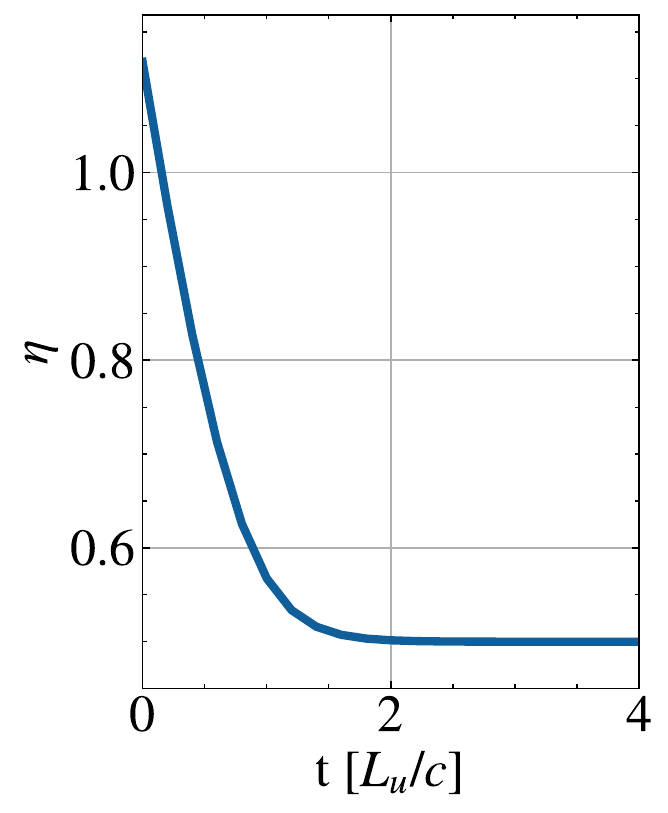}
        \caption{Same as \refig{fig:uniform_test1} but for the non-vanishing velocity gradients case.}
        \label{fig:uniform_test2}
    \end{figure}
    We further tested our implementation of the effective resistivity by extending the previous test to non-vanishing velocity gradients, i.e. $\partial_yv_y\neq0$.
    In this case Amp\`ere's law takes the form
    \begin{align} \label{eq:ampere_test2}
    	\frac {\partial \vec{E}} {\partial t} = -|J_0| \frac{\vec{E}}{\sqrt{E_0^2 + E_z^2}}\; ,
    \end{align}
    which can be reduced to a single equation for the $z$ component of the electric field as it was for \refeq{eq:ampere_test1}.
    \refeq{eq:ampere_test2} can be then integrated to obtain  
    \begin{align} \label{eq:solution_test2}
    	\sqrt{E_0^2 + E^2_z(t)} + \frac 1 2 E_0 \ln \left| \frac{\sqrt{E_0^2 + E^2_z(t)} - E_0}{\sqrt{E_0^2 + E^2_z(t)} + E_0} \right| = -|J_0| t + K\; ,
    \end{align}
    which can be solved numerically once the integration constant $K$ is set by the initial conditions.
    
    The left panel of \refig{fig:uniform_test2} shows both the field described by \refeq{eq:solution_test2} and the solution obtained by the \pluto{} code with $J_0=1$, $E_0=0.5$, and the same initial electric field as the previous test.
    Also in this case the code's calculation agrees with the analytic solution, which leads to the complete dissipation of the electric field by $t\simeq4$.
    However, its decay is no longer linear but rather exponential-like, as the non-vanishing $E_0$ introduces a minimum constant resistivity acting during the whole evolution of the system.
    This can be seen in the evolution over time of the resistivity (right panel of \refig{fig:uniform_test2}), which starts from an initial value $\eta(t=0)=\sqrt{E_0^2+E_z^2}/|J_0|\simeq1.12$ and then decays down to the terminal value $E_0/|J_0|=0.5$.
    
%%%%%%%%%%%%%%%%%%%%%%%%%%%%%%%%%%%%%%%%%%%%%%%%%%%%%%%%%%%%%%%%%%%%%%%%%%%%%%%%

    \subsection{1D self-similar current sheet}\label{app:komissarov}
    As second test for our implementation of the effective resistivity we considered the self-similar dissipation of a 1D current sheet \citep{komissarov2007}. 
    While this problem has become a standard numerical test for resistive RMHD codes, we use it here as a benchmark to characterize the action of the effective resistivity in a more realistic and physically motivated setup.

    In the limit of infinite resistivity and thermal pressure only the magnetic field evolves, with the induction equation being the only one that needs to be solved.
    In this resistive regime, the magnetic field evolves in time according to
    \begin{equation} \label{eq:komissarov}
        B_x(y, t) = B_0\text{erf} \left( \frac{y}{2 \sqrt{\eta} t} \right)\; .
    \end{equation}
    We start our simulations at the instant $t=1$ imposing $\rho=1$, $p=50$, $B_0=1$, and $\vec{B}=(B_x(y,t=1),0,0)$.
    Along with the test using the effective resistivity prescription, we also run a benchmark with constant resistivity $\eta_0=0.1$ which will serve as reference.
    The same value is used in \refeq{eq:komissarov} to set identical initial conditions for the effective resistivity case.
    The 1D computational domain covers with $N_y=4096$ grid points a transversal slice of a current sheet with $y\in[-20,20]$. 
    As for the previous tests, we use RK3 time stepping, MP5 reconstruction, and we set $|J_0|=1$.

\begin{figure}
        \centering
        \includegraphics[width=0.48\textwidth]{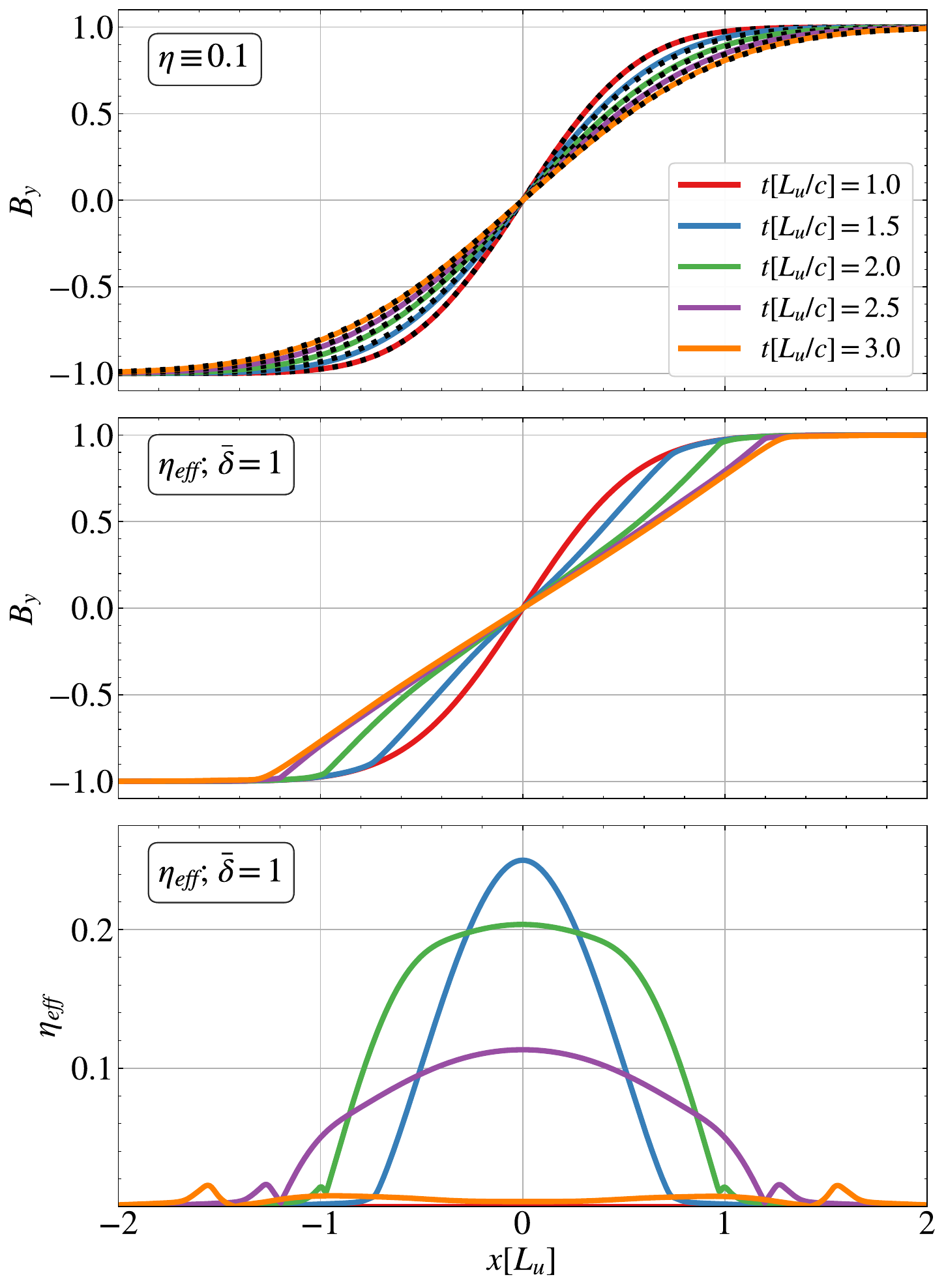}
        \caption{1D self-similar current sheet. Top panel: Spatial profiles of $B_y$ for a constant resistivity at different times.
        The analytical solution given by \refeq{eq:komissarov} is indicated with black dotted lines. 
        Mid panel: Same as the top panel, but using an effective resistivity.
        Bottom panel: Spatial profiles of effective resistivity at different times.
        }
        \label{fig:komissarov}
    \end{figure}

    The top panel of \refig{fig:komissarov} shows the evolution of the current sheet when using a constant value for the resistivity $\eta=0.1$, which reproduces the analytical profile described by \refeq{eq:komissarov} (dotted curves).
    When we use the effective resistivity prescription (mid panel) there is a localized dissipation of the current sheet closer to its center, with a continuous expanding of the region affected by the magnetic dissipation.
    The magnetic field relaxes then towards a profile that increases linearly with the domain coordinate $x$.
    This behaviour is consistent with the actual evolution of $\eta$ (bottom panel), which starts from its imposed floor value $\eta_{\rm floor}=10^{-6}$.
    It then increases to values larger than 0.1 by $t=1.5$ but only within $\Delta x\simeq1$ from the current sheet's center, whereas it remains weaker by at least two orders of magnitude for further distances.
    As time progresses, the resistivity profile becomes flatter over a larger region, while also decreasing in amplitude and approaching a mean value within the current sheet of $\sim0.01$ by $t\simeq3$. 

    \end{appendix}
\end{document}